\documentclass[fleqn,usenatbib]{mnras}

\usepackage{newtxtext,newtxmath}

\usepackage[T1]{fontenc}
\usepackage{ae,aecompl}

\usepackage{graphicx}	
\usepackage{amsmath}	
\usepackage{amssymb}
\usepackage{algorithm}
\usepackage{algorithmic}

\usepackage{times}
\usepackage{xcolor,ulem}

\usepackage{etoolbox}
\makeatletter
\patchcmd\@combinedblfloats{\box\@outputbox}{\unvbox\@outputbox}{}{\errmessage{\noexpand patch failed}}
\makeatother
\title[Porous dust grains in protoplanetary discs -- I.]{Evolution of porous dust grains in protoplanetary discs -- I. Growing grains}

\author[A. J. L. Garcia et al.]{
A. J. L. Garcia,$^{1}$\thanks{E-mail: anthony.garcia@ens-lyon.fr}
and J.-F. Gonzalez,$^{1}\thanks{E-mail: jean-francois.gonzalez@ens-lyon.fr}$
\\
$^{1}$Univ Lyon, Univ Claude Bernard Lyon 1, Ens de Lyon, CNRS, Centre de Recherche Astrophysique de Lyon UMR5574, F-69230, Saint-Genis-Laval, France\\
}

\date{Accepted 2020 February 6. Received 2020 February 6; in original form 2019 April 29}

\pubyear{2020}

\begin{document}
\label{firstpage}
\pagerange{\pageref{firstpage}--\pageref{lastpage}}
\maketitle

\begin{abstract}
One of the main problems in planet formation, hampering the growth of small dust to planetesimals, is the so-called radial-drift barrier. Pebbles of cm to dm sizes are thought to drift radially across protoplanetary discs faster than they can grow to larger sizes, and thus to be lost to the star. To overcome this barrier, drift has to be slowed down or stopped, or growth needs to be sped up. In this paper, we investigate the role of porosity on both drift and growth. We have developed a model for porosity evolution during grain growth and applied it to numerical simulations of protoplanetary discs. We find that growth is faster for porous grains, enabling them to transition to the Stokes drag regime, decouple from the gas, and survive the radial-drift barrier. Direct formation of small planetesimals from porous dust is possible over large areas of the disc.
\end{abstract}




\section{Introduction}

In protoplanetary discs, planetesimals are thought to be the building blocks of planets \citep{1972epcf.book.....S}. However, our understanding of the collisional growth of sub-micron-sized monomers up to kilometre-sized planetesimals is hindered by problems called `barriers' in the planet formation theory. 
Because of the gas pressure gradient, a differential velocity exists between gas and dust, and grains experience an aerodynamic drag force. Grains thus lose angular momentum, making them settle down to the mid-plane and drift inwards. The influence of the drag force can be measured by the Stokes number St, i.e. the ratio between the grain stopping time (the time for a grain to reach the gas velocity) to the Keplerian orbital time. Small grains, with St$\ll$1, and large grains, with St$\gg$1, are respectively very coupled with the gas and weakly affected by the gas drag and so, drift slowly. Grains with intermediate sizes, i.e. St$\sim$1, are subject to the fastest radial drift. Thus, initially small grains growing to this intermediate regime can be accreted onto the star faster than they can grow to reach planetesimal sizes \citep{1976PThPh..56.1756A,1977MNRAS.180...57W}. To overcome this barrier, called the ``radial-drift barrier'', grain growth needs to be accelerated or drift slowed down. So-called dust traps have been shown to be a good way to slow down drift by accumulating grains in a gas pressure maximum \citep{2005MNRAS.362.1015H} such as, e.g., in vortices \citep{1995A&A...295L...1B,2012A&A...545A.134M,2014ApJ...785..122Z}, snow lines \citep{2007ApJ...664L..55K,2008A&A...487L...1B,2017A&A...608A..92D} or planet gaps \citep{2004A&A...425L...9P,2007A&A...474.1037F,2010A&A...518A..16F,2012MNRAS.423.1450A,2012A&A...545A..81P,2012ApJ...755....6Z,2014ApJ...785..122Z}. More recently, \citet{2017MNRAS.467.1984G,2017MNRAS.472.1162G} showed that self-induced dust traps can form without an initially present pressure maximum in the disc, where growing and fragmenting grains can accumulate thanks to the combined effect of the back-reaction of dust on gas and large-scale gradients.

In this work, we choose to investigate on the growth of porous grains as a solution. Indeed, at a given mass, porous grains have a larger collisional cross-section than compact dust, which can accelerate growth. Until now, grains have been considered as compact in most studies of grain growth \citep[e.g.][]{2008A&A...487L...1B,2008A&A...487..265L,2014A&A...567A..38D,2015P&SS..116...48G}, for the sake of simplicity and because of a lack of knowledge on how porosity evolves with collisions. However, some objects in the Solar system such as comets have appeared to be porous \citep{2006ApJ...652.1768B,2011ARA&A..49..281A,2015Sci...347a1044S}. Experiments have also shown that low-velocity collisions lead to the formation of porous aggregates \citep{2004ASPC..309..369B}. Furthermore, experimental and numerical works have studied faster collisions and found that collisional energy is dissipated by internal restructuring and that compression can occur \citep{1997ApJ...480..647D,2000Icar..143..138B,2007ApJ...661..320W,2008ApJ...677.1296W,2009ApJ...702.1490W,2008ApJ...684.1310S,2012ApJ...753..115S,2012A&A...541A..59S}. The importance of porosity on the collisional evolution of dust grains, and in particular its ability to assist growth, was first evidenced in numerical simulations by \citet{2007A&A...461..215O}. Based on the numerical model of evolution of porosity during collisions of \citet{2008ApJ...684.1310S}, \citet{2012ApJ...752..106O} then investigated the formation of icy planetesimals from direct growth of porous aggregates and found that porosity can help grains overcome the radial-drift barrier. However, in their study, planetesimals have a density lower than $10^{-3}$~kg\,m$^{-3}$, inconsistent with comet values. \citet{2013A&A...557L...4K} thus introduced the static compression process to take into account restructuring due to gas or self-gravity compression. As it is difficult and impractical to treat at the same time the evolution of porosity during collisions at small scales and the disc physics at large scales, studies including porosity have mostly been carried out in a local context, i.e. at a given distance from the star \citep{2007A&A...461..215O,2009ApJ...707.1247O,2013A&A...557L...4K}. However, \citet{2012ApJ...752..106O} and \citet{2016A&A...586A..20K} computed the radial evolution of grain mass and porosity in a 1D disc model. In this paper, in order to link both the small and large scales, we derive a model of porosity evolution to reproduce the effects of collisions and include it in global 3D simulations of dust growth and dynamics.

We detail our models for growth and porosity as well as our numerical codes in Section~\ref{Sec2}. In Section~\ref{Sec3}, we present our results on the influence of porosity on growth and drift and we discuss on how porosity can be a solution to the radial-drift barrier in Section~\ref{Sec4}. We summarise our work and conclude in Section~\ref{Sec5}.

\section{Methods}
\label{Sec2}
\subsection{Disc models}
\label{sec:disc}

We simulate a disc of mass $M_\mathrm{disc} = 0.01\ M_\odot$ around a 1~$M_\odot$ star. Initially, the dust-to-gas ratio by mass $\epsilon$, defined as the ratio between dust and gas local volume densities $\rho_\mathrm{d}/\rho_\mathrm{g}$, is taken to be 0.01. We model the disc as power laws of the distance from the star $R$. Thus, the gas surface density scales as $\Sigma_\mathrm{g} \propto R^{-p}$ and its temperature as $T_\mathrm{g} \propto R^{-q}$. In all simulations, we use a constant \citet{1973A&A....24..337S} $\alpha$ parameter of $10^{-2}$. The disc extends from $R_\mathrm{in}$ to $R_\mathrm{out}$. We assume that dust moving outside of a distance $R_\mathrm{esc}$ or inside $R_\mathrm{in}$ is respectively escaped or accreted into the star.

\begin{table}
\centering
\caption{Parameters of our disc models. Values with subscript 0 are given for $R_\mathrm{0}$ = 1 AU.}
\label{Tableau1}
\begin{tabular}
{llllllll}
\hline
Model & $p$ & $q$ & $R_{\mathrm{in}}$ & $R_{\mathrm{out}}$ & $R_{\mathrm{esc}}$ & $T_\mathrm{g,0}$ & $\Sigma_{\mathrm{g,0}}$ \\
	& & & (AU) & (AU) & (AU) & (K) & (kg.m$^{-2}$) \\
\hline\hline
CTTS & 3/2 & 3/4 & 3 & 300 & 400 & 197 & 4537 \\
\hline
Flat & 0 & 1 & 4 & 120 & 160 & 619 & 11 \\
\hline
\end{tabular}
\end{table}

We consider two disc models characterised by two different disc geometries: a disc around a classical T-Tauri star (CTTS) \citep[studied by, e.g.,][]{2005A&A...443..185B,2008A&A...487..265L} and a ``Flat'' disc, a simple model with a constant surface density \citep[used in, e.g.,][]{2004A&A...425L...9P,2007A&A...474.1037F,2015P&SS..116...48G}. Both models correspond to different outcomes regarding the radial-drift barrier for non-growing grains according to \citet{2012A&A...537A..61L}: dust grains are expected to survive the barrier in the CTTS disc but not in the Flat disc. We study both in this work to examine whether porosity impacts this behaviour. The different parameters of those disc are given in Table~\ref{Tableau1}.

\subsection{Drag regimes and drift}
\label{sec:DR}

The differential velocity between gas and dust creates an aerodynamic friction. Grains are in different drag regimes according to how their size $s$ compares to the mean free path of gas molecules
$\lambda$. \citet{1996A&A...309..301S} give the expression of the gas dynamical viscosity as a function of $\lambda$ as
\begin{equation}
\mu_\mathrm{g}=\frac{1}{2}\,\rho_\mathrm{g} \, \lambda \, c_\mathrm{g},
\end{equation}
where
\begin{equation}
c_\mathrm{g}=\sqrt{\frac{k_\mathrm{B}\,  T_\mathrm{g}}{m_\mathrm{g}}}
\end{equation}
is the gas sound speed, with $m_\mathrm{g}$ the mean gas molecule mass. This yields
\begin{equation}
\label{eq:mfp}
\lambda=\frac{2\mu_\mathrm{g}}{\rho_\mathrm{g} \, c_\mathrm{g}}.
\end{equation}
The Chapman-Enskog theory provides a refined calculation of $\mu_\mathrm{g}$. In the Sutherland model, which describes rigid elastic spheres with weak mutual attraction, it can be approximated as
\begin{equation}
\mu_\mathrm{g} = \frac{5\sqrt{\pi} \, m_\mathrm{g} \, c_\mathrm{g}}{64 \, \sigma_\mathrm{mol}}
\end{equation}
\citep{1970mtnu.book.....C}, where $\sigma_\mathrm{mol} = 2\times10^{-19}$~m$^2$ is the cross-section of the H$_2$ molecule, and we take $m_\mathrm{g} = 2.32\ m_\mathrm{H} = 3.883038752\times10^{-27}$~kg.
When $s<9\lambda/4$, the grain is in the Epstein drag regime \citep{1924PhRv...23..710E} and when $s>9\lambda/4$, it is in the Stokes regime \citep{1972fpp..conf..211W}. In those regimes, the grain's Stokes number can be expressed as
\begin{equation}
\label{Stokesnumber}
\mathrm{St} =
\left\{ \begin{array}{ll}
\displaystyle\frac{\rho_\mathrm{s} \, \phi \, s}{\rho_\mathrm{g} \, c_\mathrm{g}}\Omega_\mathrm{K}\,,   \qquad & (\mathrm{Epstein})\\
  \\
\displaystyle\frac{2 \, \rho_\mathrm{s} \, \phi \, s^2}{9 \, \mu_\mathrm{g}}\Omega_\mathrm{K} \,, \qquad  &(\mathrm{Stokes}) 
   \end{array}
\right.  
\end{equation}
where $\rho_\mathrm{s}$ is the bulk density of the grain and $\Omega_\mathrm{K}$ is the Keplerian angular velocity. $\phi$ is the grain filling factor and will be discussed in Section~\ref{sec:poro}. Note that some authors use the gas sound speed while others use the mean thermal velocity of gas molecules in the calculation of the drag force in the Epstein regime. We use the former. The resulting expressions for St differ by a factor $\sqrt{\pi/8}$, of order unity. In the Stokes regime, even though the Stokes number would depend on the Reynolds number Re \citep{1972fpp..conf..211W}, we deliberately do not use the Re dependency of the Stokes number for large Re for the sake of simplicity as it would require iterations to calculate it (Re depends on the differential velocity between gas and dust, which in turn depends on the stopping time, which is a function of Re). This amounts to limiting ourselves to the linear Stokes drag. Similarly, \citet{2012ApJ...752..106O} neglected the high-Re domain of the Stokes regime (also known as the Newton drag regime) for simplicity. They showed that this somewhat accelerates dust growth but concluded that it has little effect of the ability of porosity to help grains to survive the radial-drift barrier. Additionally, the transition to the Newton regime would only occur in the inner disc regions for bodies larger than $\sim100$~m, for which self-gravity would start to be important and assist their growth. However, this is out of the scope of this paper (see Section~\ref{sec:SPH}).

This friction force makes the dust settle down to the mid-plane and drift towards the star. \citet{1986Icar...67..375N} gave the dust radial velocity as
\begin{equation}
\label{vdre}
v_{\mathrm{d},R}=-\frac{\mathrm{St}}{(1 + \epsilon)^2 + \mathrm{St}^2}\,\eta \,v_\mathrm{K}  \,,
\end{equation}
where $v_\mathrm{K}$ is the Keplerian velocity and $\eta$ the sub-Keplerian parameter defined as
\begin{equation}
\label{eta}
\eta= - \left(\frac{H_\mathrm{g}}{R}\right)^2 \frac{\partial \ln P_\mathrm{g}}{\partial \ln R} \,,
\end{equation}
with $H_\mathrm{g}$ the gas scale height and $P_\mathrm{g}$ the gas pressure. From equation~(\ref{vdre}), we can find again that both small and large grains, with small and large St values, drift slowly. We can also infer that dust can be slowed down by increasing the dust-to-gas ratio $\epsilon$. 

\subsection{Growth model}
\label{sec:GM}

In this study, we only consider icy dust with $\rho_\mathrm{s}=917$~kg\,m$^{-3}$. Indeed, water ice can be found in large quantities exterior to the snow line \citep{2014prpl.conf..835V}. Furthermore, in those regions, water can condense on grains surface, change the sticking properties and promote growth over bouncing or fragmentation \citep{2011Icar..214..717G,2015ApJ...798...34G}. Silicate grains have a fragmentation threshold $\sim1$~m\,s$^{-1}$ while icy dust fragments for a relative velocity larger than several 10~m\,s$^{-1}$ \citep{2008ARA&A..46...21B,2014ApJ...783L..36Y}. While recent laboratory experiments \citep{2018MNRAS.479.1273G,2019ApJ...873...58M} found lower surface energies for water ice than previously used at low temperatures, thus suggesting that ice is not more resistant than silicates, they disagree with the tensile strengths measured in numerical simulations by \citet{2019ApJ...874..159T}. Further investigation is thus needed before the issue can be settled. Thus, in this work we focus on pure growth and neglect bouncing and fragmentation, similarly to \citet{2016A&A...586A..20K}. This also allows to better understand the influence of porosity on these physical processes separately. The actual value of the fragmentation threshold of water ice therefore does not affect the results presented here. Bouncing and fragmentation will be taken into account in a followup paper.

We use a grain growth model with a locally mono-disperse mass distribution, in which collisions are considered to occur between identical grains. Details on the model implementation and a discussion on its assumptions can be found in \citet{2008A&A...487..265L}. As fragmentation is not taken into account in this study, every collision leads to sticking. \citet{1997A&A...319.1007S} gave the expression of the variation of a grain's mass with time as it doubles in a mean collision time $\tau_\mathrm{coll}$, i.e. 
\begin{equation}
\label{dmdt1}
\frac{\mathrm{d}m}{\mathrm{d}t}\approx\frac{m}{\tau_\mathrm{coll}}= 4 \pi s^2 \, v_\mathrm{rel} \, \rho_\mathrm{d} = 4 \pi s^2 \, v_\mathrm{rel} \, \epsilon\, \rho_\mathrm{g} \:. 
\end{equation}
Grains collide with a relative velocity $v_\mathrm{rel}$ transmitted from gas turbulent motion to the dust by aerodynamic drag. We use the \citet{1997A&A...319.1007S} model, in which
\begin{equation}
\label{vrel1}
v_\mathrm{rel} = \sqrt{2^{3/2}\, \mathrm{Ro}\, \alpha}\, \frac{\sqrt{\mathrm{Sc}-1}}{\mathrm{Sc}} \, c_\mathrm{g},
\end{equation} 
where Ro is the Rossby number for turbulent motions, taken equal to 3, and Sc the Schmidt number of the grains, whose expression is
\begin{equation}
\label{schmidt}
\mathrm{Sc} = (1+\mathrm{St}) \, \sqrt{1 + \frac{\Delta v^2}{v_\mathrm{t}^2}},
\end{equation} 
where $v_\mathrm{t}=\sqrt{2^{1/2}\, \mathrm{Ro}\,\alpha} \, c_\mathrm{g}$ denotes the turbulent velocity. $\Delta v = v_\mathrm{d} - v_\mathrm{g}$ is the differential velocity between dust and gas. In our simulations, $\Delta v$ appeared to be negligible compared to $v_\mathrm{t}$ \citep[see also][]{2008A&A...487..265L} and $v_\mathrm{rel}$ can be approximated as
\begin{equation}
\label{vrel2}
v_\mathrm{rel} \simeq \sqrt{2^{3/2}\, \mathrm{Ro}\, \alpha} \, \frac{\sqrt[]{\mathrm{St}}}{1+\mathrm{St}} \: c_\mathrm{g}.
\end{equation}
The reader is referred to \citet{2008A&A...487..265L} for a discussion of the difference between equation~(\ref{schmidt}) and the expression of \citet{2007Icar..192..588Y}, and to \citet{2014MNRAS.437.3037L} for a more general discussion on various models for relative velocities.

The collisional kinetic energy $E_\mathrm{kin}$ for identical grains is then given by
\begin{equation}
\label{Ekin}
E_\mathrm{kin} = \frac{1}{2} \, m^* \, v_\mathrm{rel}^2 = \frac{1}{4} \, m \, v_\mathrm{rel}^2\,,
\end{equation}
where the reduced mass $m^*=m/2$ for identical grains.

\subsection{Model for porosity evolution during growth}
\label{sec:poro}

We consider grains as aggregates of elementary monomers. Those monomers are compact spheres, with a density $\rho_\mathrm{s}$ and a radius $a_0$. The volume filling factor $\phi$ is then used to characterise how porous the grains are. It is defined as the ratio between the volume occupied by matter $V_\mathrm{mat}$ to the total volume $V$ of a grain:
\begin{equation}
\label{defphi}
\phi = \frac{V_\mathrm{mat}}{V} = \frac{\rho}{\rho_\mathrm{s}}
\end{equation}
where $\rho$ is the density of the grain. Compact grains have $\phi = 1$ while $\phi \sim 0$ corresponds to very fluffy aggregates.

Real porous aggregates can have arbitrary shapes and mass distributions. Parameters such as radius, cross-section or volume are therefore ill-defined. The equations presented in Sections~\ref{sec:DR} and \ref{sec:GM} were initially derived for compact spherical grains. In this work, in order to be able to use them, as well as equation~(\ref{defphi}), we assume that a porous aggregate can be represented by an equivalent spherically-symmetric and non-fractal collection of monomers in order to describe its spatial, size and porosity evolution.

\subsubsection{Collisional evolution}
\label{sec:colevol}

Small grains, with low velocities \citep{1993prpl.conf.1031W} meet slowly and stick, it is the ``hit-and-stick'' regime. Such collisions make the grains more porous, or decrease their filling factor, as their total volume grows faster than that occupied by matter \citep[see][]{2012ApJ...752..106O}. When grains collide with a kinetic energy larger than the rolling energy $E_\mathrm{roll}$ (see equation~\ref{eq:Eroll}), i.e.\ the energy needed for one monomer to roll over 90$^{\circ}$ on the surface of another monomer \citep{1997ApJ...480..647D}, the formed aggregate dissipates the extra energy by compressing. Thus, the grain still becomes more porous, but less than in the hit-and-stick regime \citep{2012ApJ...752..106O}. Thanks to numerical simulations of collisions, \citet{2008ApJ...684.1310S} have expressed the filling factor of the grain after a collision $\phi_\mathrm{f}$ in those two regimes as a function of parameters before the collision such as the kinetic energy $E_\mathrm{kin}$ or the filling factor $\phi_\mathrm{i}$ of colliding grains. However, this formula is recursive and cannot be used in our simulations where a time step corresponds to any number (not necessarily an integer) of collisions. Thus, to describe the evolution of the filling factor during collisional growth $\phi_\mathrm{col}$, we needed to modify the \citet{2008ApJ...684.1310S} model to make it continuous and non-recursive. This amounts to obtaining an expression for $\phi_\mathrm{f}$ only as a function of grain mass $m$ and disc quantities. The details of our calculations and approximations are presented in Appendix~\ref{App1}. Our resulting expressions for $\phi_\mathrm{col}$ are:
\begin{enumerate}
\item hit-and-stick
\begin{equation}
\label{eq:phiHS}
\phi_{\mathrm{h\&s}} = \left(\frac{m}{m_\mathrm{0}}\right)^{-0.58} ,
\end{equation}
where $m_\mathrm{0}=\frac{4}{3}\pi a_0^3\,\rho_\mathrm{s}$ is the monomer mass.
\item collisional compression: four cases need to be considered depending on the drag regime, Epstein or Stokes, and whether $\mathrm{St}<1$ or $\mathrm{St}>1$
\begin{align}
\label{phiEpSt<1}
&\phi_\mathrm{Ep-St<1} = \left(2^{1/5} - \beta_\mathrm{Ep}\right)^{-3/8} \, \beta_\mathrm{Ep}^{-5/8} \, 2^{1/8} \nonumber\\
&\quad\times \left(\frac{3}{10} \frac{(3/5)^5 \, 2^{3/2} \, \mathrm{Ro}\, \alpha \, m_\mathrm{0} \, c_\mathrm{g}\, \rho_\mathrm{s} \, a_\mathrm{0} \, \Omega_\mathrm{K}}{8\,\rho_\mathrm{g}\,b\,E_\mathrm{roll}}\right)^{3/8} \left(\frac{m}{m_\mathrm{0}}\right)^{-1/8} \,, 
\end{align}
\begin{align}
\label{phiStSt<1}
&\phi_\mathrm{St-St<1} = \left(2^{1/5} - \beta_\mathrm{St}\right)^{-1/3}\, \beta_\mathrm{St}^{-2/3} \nonumber\\
&\quad\times \left(\frac{3}{10} \frac{(3/5)^5 \, 2^{3/2} \, \mathrm{Ro}\, \alpha \, m_\mathrm{0} \, c_\mathrm{g}^2\, \rho_\mathrm{s} \, a_\mathrm{0}^2 \,\Omega_\mathrm{K}}{36\,\mu_\mathrm{g}\,b\,E_\mathrm{roll}}\right)^{1/3} , 
\end{align}
\begin{align}
\label{phiEpSt>12}
\phi_\mathrm{Ep-St>1} = \phi_\mathrm{Ep-St<1}(M_\mathrm{4}) \, \left(\frac{m}{M_\mathrm{4}}\right)^{-1/5} \,,
\end{align}
\begin{align}
\label{phiStSt>12}
\phi_\mathrm{St-St>1} = \phi_\mathrm{St-St<1}(M_\mathrm{5}) \, \left(\frac{m}{M_\mathrm{5}}\right)^{-1/5} \,,
\end{align}
where $M_\mathrm{4}$ and $M_\mathrm{5}$ are the masses for which $\mathrm{St} = 1$, respectively in the Epstein and Stokes regimes. They are given in equations~(\ref{M4}) and (\ref{M5}).
\end{enumerate}

\subsubsection{Static compression}
\label{sec:static}

\citet{2013A&A...554A...4K} have shown that very fluffy grains can be statically compressed by an applied compressive strength $P$. The filling factor is then linked to $P$ as
\begin{equation}
\label{phikataoka}
\phi = \left(\frac{a_0^{3}}{E_\mathrm{roll}} \, P \right)^{1/3}\, .
\end{equation}
Thus, the gas drag force $F_\mathrm{drag}$ applies a compressive strength $P_\mathrm{drag} = F_\mathrm{drag}/(\pi \: s^2)$ that can compact the grain up to a value $\phi_\mathrm{gas}$ depending on the drag regime:
\begin{enumerate}
\setcounter{enumi}{2}
\item gas drag compression
\begin{equation}
\label{phigasEp}
\phi_\mathrm{gas} = \left\{
    \begin{array}{ll}
      \displaystyle\left(\frac{m_\mathrm{0} \, \Delta v}{\pi \, E_\mathrm{roll}}\frac{\rho_\mathrm{g} \, c_\mathrm{g}}{\rho_\mathrm{s}}\right)^{1/3} \,, & \mbox{(Epstein)} \\ 
 \\
      \displaystyle\left(\frac{6 \, a_\mathrm{0}^2 \, \Delta v \, \mu_\mathrm{g}}{E_\mathrm{roll}}\right)^{3/8} \left(\frac{m}{m_\mathrm{0}}\right)^{-1/8}  \,.& \mbox{(Stokes)}
    \end{array}
\right.
\end{equation}
\end{enumerate}
Finally, massive enough grains can be compressed by their self-gravity up to a filling factor $\phi_\mathrm{grav}$ given by \citet{2013A&A...557L...4K} as:
\begin{enumerate}
\setcounter{enumi}{3}
\item self-gravity compression
\begin{equation}
\label{phigrav}
\phi_\mathrm{grav}= \left(\frac{\mathcal{G} \, m_\mathrm{0}^2}{\pi \, a_\mathrm{0} \, E_\mathrm{roll}}\right)^{3/5} \left(\frac{m}{m_\mathrm{0}}\right)^{2/5} \,.
\end{equation}
\end{enumerate}

The filling factor of dust naturally evolves as $\phi_\mathrm{col}$ during collisions until the grain is fluffy enough to be compressed by the gas drag or its self-gravity and then its filling factor is equal to $\phi_\mathrm{gas}$ or $\phi_\mathrm{grav}$ respectively. Therefore, the maximum of $\phi_\mathrm{col}$, $\phi_\mathrm{gas}$ and $\phi_\mathrm{grav}$ 
provides the smallest possible value of the filling factor. The different regimes encountered by growing grains at fixed locations are summarised in Figure~\ref{phimana2}.

\subsection{Numerical simulations}
\label{sec:HS} 

In order to model the evolution of porous dust grains in protoplanetary discs, we use two codes with different properties.

\subsubsection{The PACED code}
\label{sec:PACED}

We have developed the PACED\footnote{for Porous And Compact dust Evolution in Discs} code, a 1D C++ code to model the behaviour of a single grain at a time in a stationary gas disc described with power laws with a constant dust-to-gas ratio of $10^{-2}$. This code contains radial drift with equation~(\ref{vdre}), the growth model described in Section~\ref{sec:GM} and porosity evolution with the model presented in Section~\ref{sec:poro}.

\subsubsection{3D SPH code}
\label{sec:SPH}

We also use the 3D, two-phase (gas+dust), SPH (Smoothed Particle Hydrodynamics) code described in \citet{2005A&A...443..185B}. The code computes the dynamics of both phases, which interact via aerodynamic drag in the Epstein regime, including the back-reaction of dust on gas. (In contrast to PACED, this code does not use equation~(\ref{vdre}) but calculates the forces on each SPH particle and their ensuing motion.) It reproduces the properties of turbulence \citep{2013MNRAS.433...98A}. Grain growth was implemented in \citet{2008A&A...487..265L} and 
fragmentation in \citet{2015P&SS..116...48G}. Self-gravity is not included. Detailed discussions of the code properties can be found in \citet{2007A&A...474.1037F} and \cite{2017MNRAS.467.1984G}. In this work, we have added the implementation of the Stokes drag regime to consider the evolution of the largest solids. We have also included our model of porosity evolution.

We start the simulations presented here with 200,000 SPH gas particles arranged to reproduce the power laws presented in Section~\ref{sec:disc} and let the disc relax for 24 orbits at $R=100$~AU for the CTTS disc and $R=40$~AU in the Flat disc. Then, the same number of SPH dust particles is injected at the same locations and velocities as the gas particles, to recreate an initial dust-to-gas ratio $\epsilon=10^{-2}$. Both sets of particles are then evolved together. Previous work with the same code \citep[e.g.][]{2005A&A...443..185B, 2007A&A...474.1037F,2019MNRAS.490.4428P} has shown that convergence is reached at lower particle numbers. We also verify that the resolution criterion from \citet{2012MNRAS.420.2345L} is met. Each SPH dust particle represents a collection of identical physical dust grains (locally mono-disperse assumption, see Section~\ref{sec:GM}) and carries intrinsic dust properties such as mass, filling factor and thus, size, that are described according to the models presented in Sections~\ref{sec:GM} and \ref{sec:poro}. In particular, we chose a uniform initial size $s_0=10\ \mu$m for all grains to shorten the computation times. Indeed, we have previously shown that very small grains grow fast and quickly forget their initial size \citep{2008A&A...487..265L}. Their corresponding initial filling factor and mass, which depend on their location, are pre-calculated with the PACED code by evolving compact monomers of size $a_0$ without radial drift. The simulations are stopped when the largest grains reach kilometre sizes, for which self-gravity becomes important, because both our growth model and our SPH code do not take it into account. Obtaining estimates of the fraction of dust that would turn into planetesimals or of properties of the resulting planetesimal population is therefore not feasible in the framework of our study.

\subsubsection{Comparison}
\label{sec:comparison}

\begin{table}
\centering
\caption{Comparison between our PACED and 3D SPH codes.}
\label{Tableau2}
\begin{tabular}{lcc}
\hline
Characteristics & PACED & 3D SPH \\
\hline
\hline
dynamics & X & X\\
\hline
growth & X & X \\
\hline
porosity & X & X\\
\hline
turbulence & & X\\
\hline
disc evolution & & X\\
\hline
collective effects & & X\\
\hline
\end{tabular}
\end{table}

As summarised in Table~\ref{Tableau2} and contrary to the PACED code, our 3D SPH code models the evolution of the gas disc and takes into account the collective effects. Indeed, dust drift can change the local dust-to-gas ratio which has an impact on dust growth and dynamics, while it is taken constant in the PACED code. However, the 3D SPH code requires long calculation times and is not practical to study a large variety of cases. Thus, we use the PACED code to test our model and estimate the dust behaviour on larger time scales. Moreover, in the PACED code, disc quantities such as gas density or differential velocity between gas and dust are obtained from analytical prescriptions while in the 3D SPH code, they are computed self-consistently \citep{2005A&A...443..185B}.

\section{Results}
\label{Sec3}

In Sections~\ref{subsec1} and \ref{subsec2}, we present how porosity evolves during dust growth and how it impacts growth and dynamics in the CTTS disc with monomer size $a_\mathrm{0} = 0.1\ \mu$m. In Section~\ref{subsec3}, we show the influence of the monomer size $a_\mathrm{0}$ on porosity and thus, on growth in the Flat disc.

\subsection{Evolution of the filling factor}
\label{subsec1}

\begin{figure}
\centering
\includegraphics[width=\columnwidth]{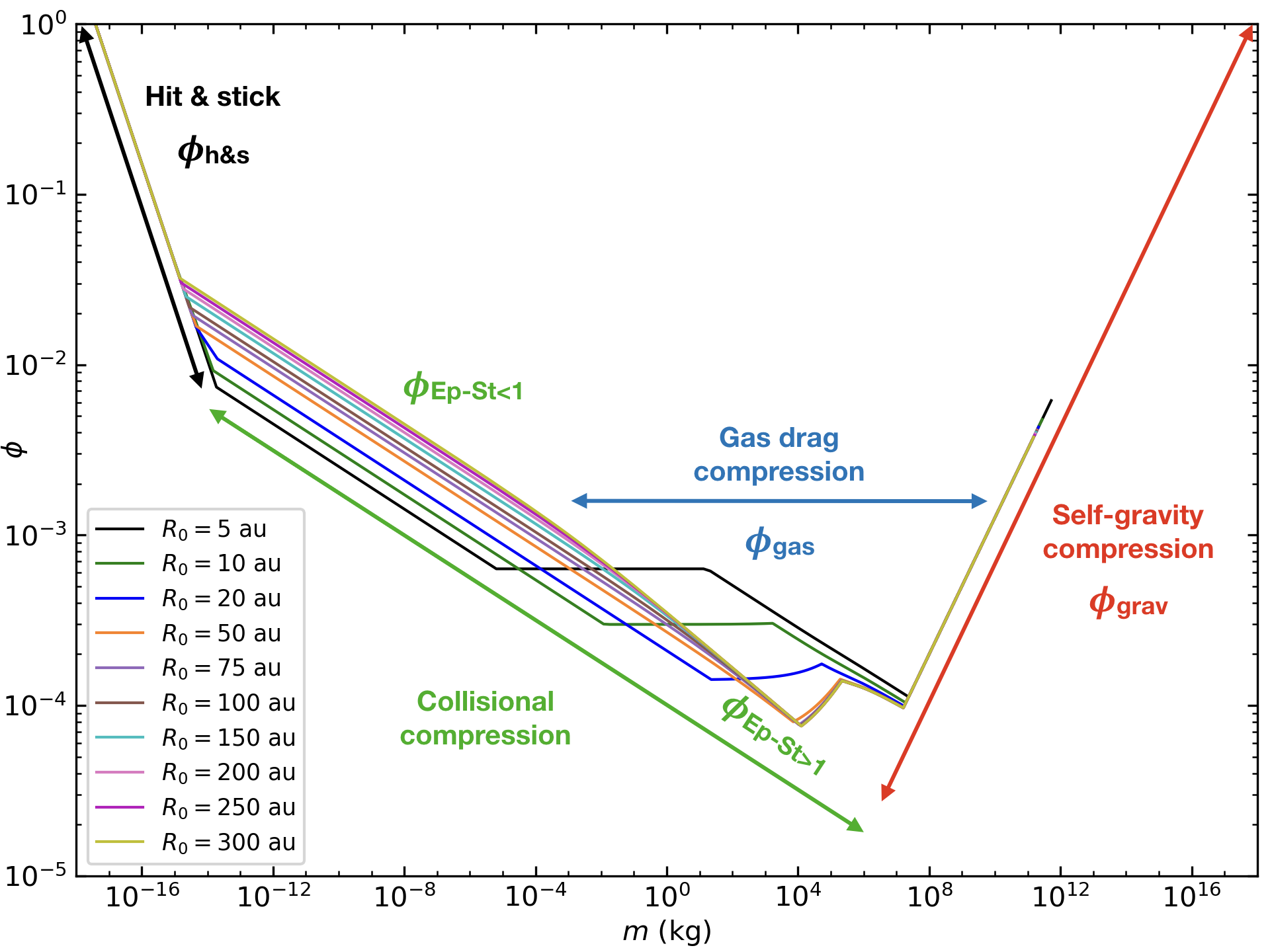}
\caption{Filling factor as a function of mass in the CTTS disc for drifting grains of different initial positions $R_\mathrm{0}$ with the PACED code. Arrows show the different regimes encountered by the grains.}
\label{phimana}
\end{figure}

Figure~\ref{phimana} shows the evolution of the filling factor of grains of different initial positions as a function of their mass as they grow and drift radially, computed with the PACED code. The curves have similar shapes to those found by \citet{2013A&A...557L...4K} for grains at fixed distances from the star (see also our model at fixed distances in Fig.~\ref{phimana2} for comparison): $\phi$ first decreases due to collisions, then increases again with gas drag compression before self-gravity is strong enough to compact the grains further. However, inwards drifting grains experience different disc conditions, and in particular larger gas densities, than static ones and are all the more compacted by gas drag as they are close to the star. Their minimum filling factor, $\sim10^4$, is somewhat larger and it is reached for smaller masses.

\begin{figure}
\centering
\includegraphics[width=\columnwidth]{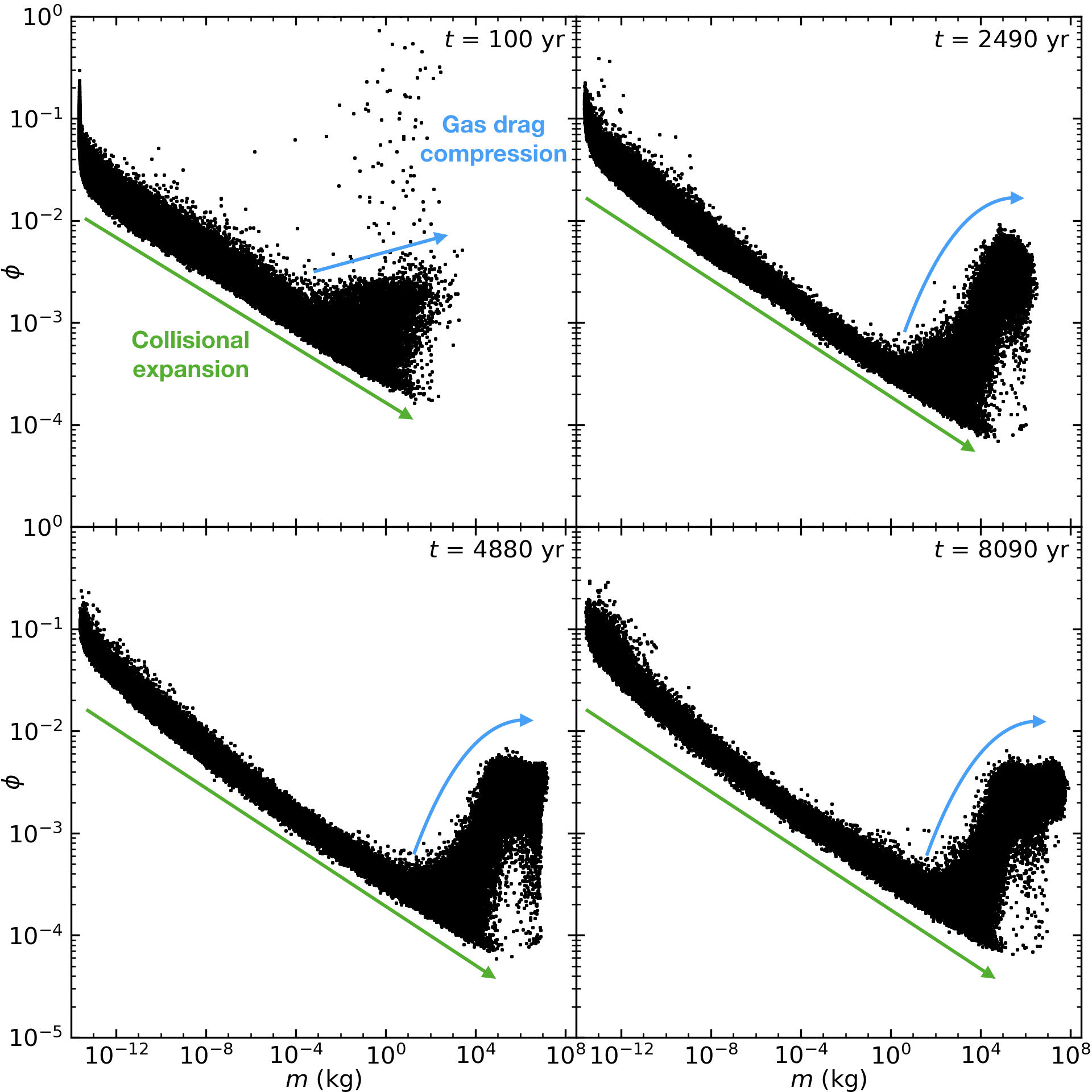}
\caption{Time evolution of the filling factor as a function of mass for grains in the CTTS disc with the 3D SPH code. Representative dots move to the right as grains grow. The filling factor decreases thanks to the collisional expansion regime (green arrow) then increases because of the compression due to the gas drag (blue arrow).}    
\label{phimCTTS}
\end{figure}

The time evolution of the grain filling factor computed with the 3D SPH code is displayed in Fig.~\ref{phimCTTS}. Initially, 10-$\mu$m sized grains have a filling factor of $\sim 10^{-1}$ -- $10^{-2}$, according to their initial location in the disc. As dust grows, representative dots move to the right. At first, grains are in the collisional compression regime. This regime tends to make grains more and more porous until they are fluffy enough to be compressed by gas drag. Note that the gas drag compression regime has a different shape than that computed with the PACED code (Fig.~\ref{phimana}), illustrating the differences between both codes listed in Section~\ref{sec:comparison}. In the top left panel, some dots move upwards with a quasi constant mass. Those dots correspond to grains that are quickly compacted by gas drag as they are accreted into the central star. The minimum value of $\phi$, also $\sim10^4$, is consistent with the numerical simulations of collisions of \citet{1999Icar..141..388K}.

\begin{figure}
\centering
\includegraphics[width=\columnwidth]{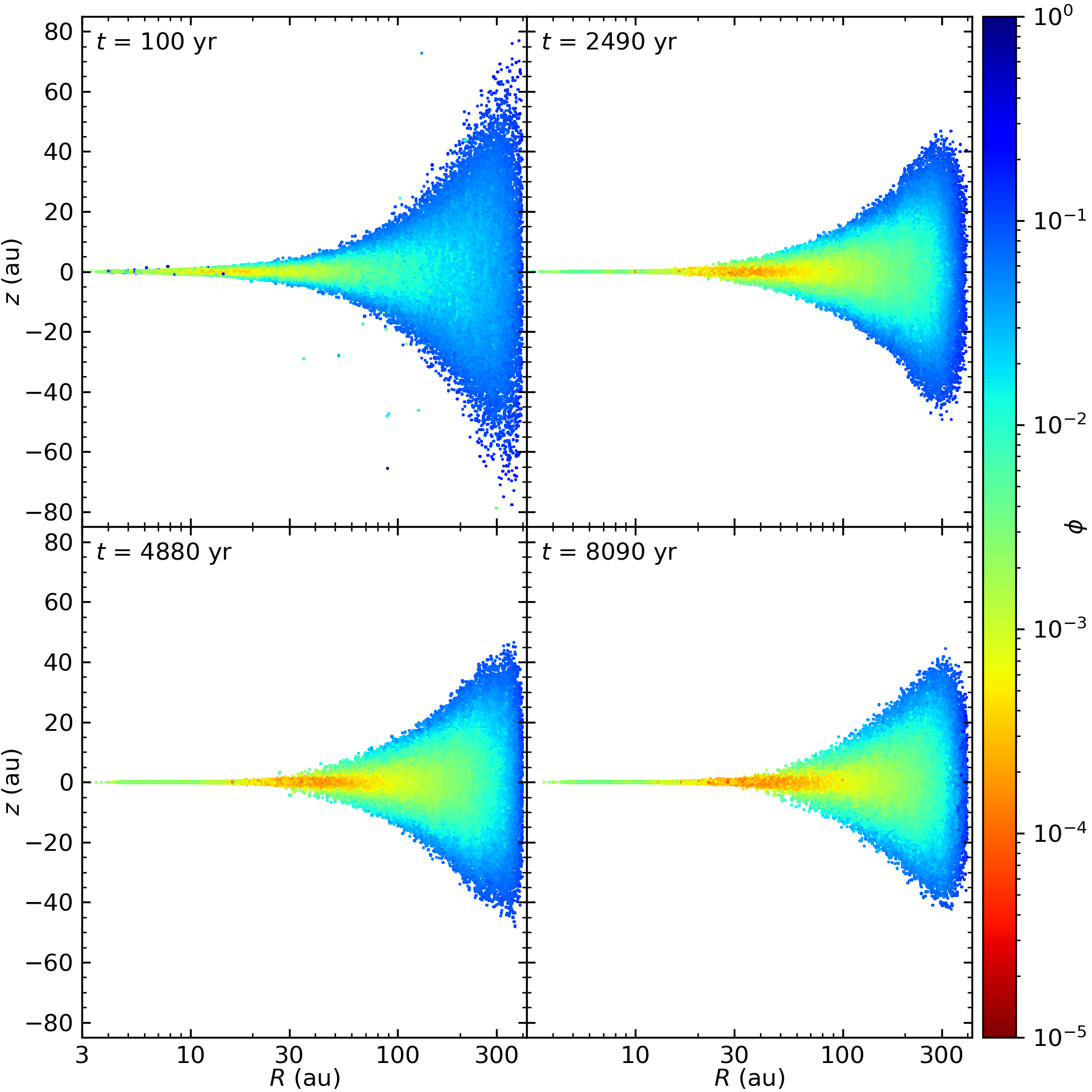}
\caption{Time evolution of the dust distribution in the CTTS disc with the 3D SPH code. The colour bar represents the filling factor.}    
\label{zRphiCTTS}
\end{figure}

Figure~\ref{zRphiCTTS} shows the distribution of the filling factor through the disc. A vertical sorting of the filling factor occurs: more compact grains are located along the disc surface while grains are fluffier and fluffier as they are closer to the mid-plane. Indeed, grains grow more rapidly in denser regions (equation~\ref{dmdt1}) and thus reach lower filling factors in the collisional regime. The same effect is seen in the radial direction. However, grains in the disc innermost regions are compacted because it is there that gas drag compression is mostly effective, where the gas is dense enough.

\subsection{Influence on grain size and dynamics}
\label{subsec2}

\begin{figure}
\centering
\includegraphics[width=\columnwidth]{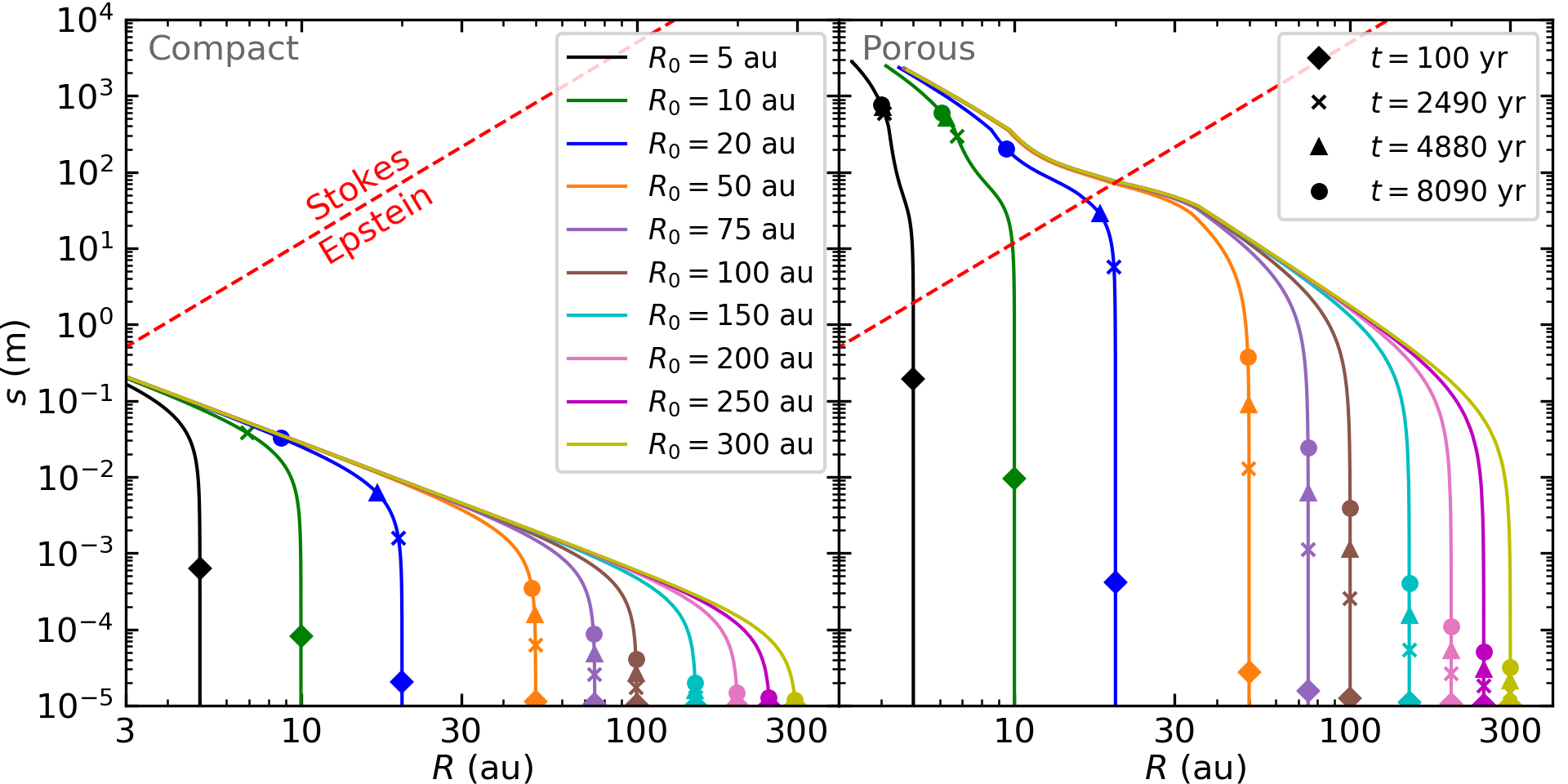}
\caption{Grain size as a function of the distance from the star in the CTTS disc for different initial positions $R_\mathrm{0}$ with the PACED code, for compact (left) and porous (right) dust. The red dashed line marks the limit between the Epstein and Stokes regimes. The symbols show the value of the size at the same time steps as in Fig.\ref{sRphiCTTS}.}
\label{sRCTTS}
\end{figure}

Figure~\ref{sRCTTS} compares the radial evolution of the size of compact and porous grains, computed with the PACED code. Compact and porous grains both have a qualitatively similar behaviour: they first grow at a quasi-constant distance from the star until they approach an optimal size $s_{\mathrm{St}=1}$, for which $\mathrm{St}=1$ and radial drift is the fastest. They then drift rapidly inwards while continuing to grow before slowing down close to the star, where they finally keep growing while drifting very little. These three stages were described and explained by \citet{2008A&A...487..265L}. During the second stage, compact grains gain 1 to 3 orders of magnitude in size before reaching the disc inner regions, depending on their initial location, while porous grains can gain up to 4. Indeed, since $s_{\mathrm{St}=1}$ is larger for porous grains (see Fig.~\ref{sSt1R}), they have a larger cross section and grow more rapidly (equation~\ref{dmdt1}): the slope in Fig.~\ref{sRCTTS} is steepest. These results are in agreement with those found by \citet{2012ApJ...752..106O}.

\begin{figure}
\centering
\includegraphics[width=\columnwidth]{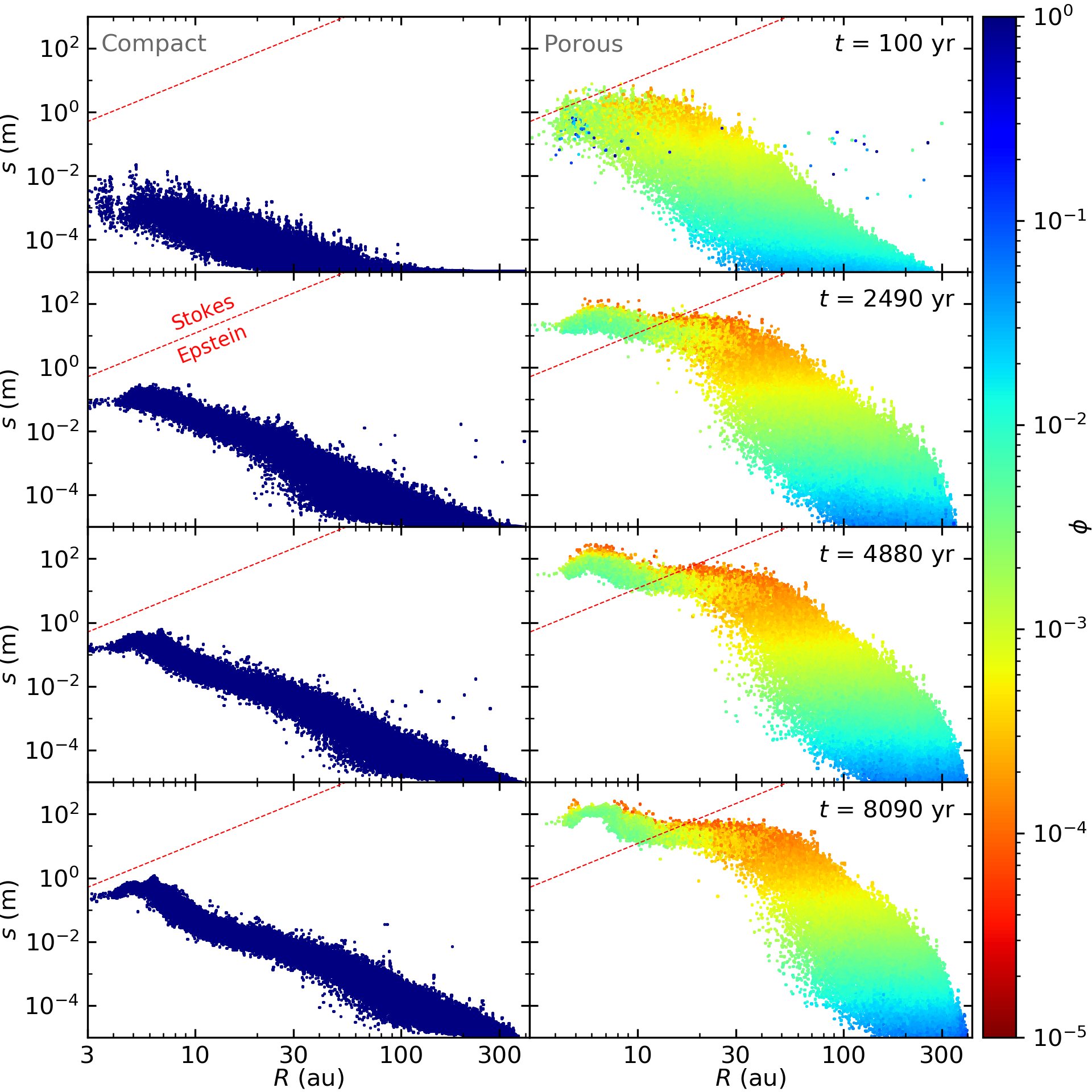}
\caption{Radial grain size distribution in the CTTS disc obtained with the 3D SPH code. Porous grains (right) can reach larger sizes than compact dust (left). The colour bar represents the filling factor. The red dashed line in the top left corner separates the Epstein (below the line) and Stokes (above the line) drag regimes. Four snapshots at 100, 2490, 4880 and 8090 yr are shown, from top to bottom.}    
\label{sRphiCTTS}
\end{figure}

The time evolution of the radial grain size distribution of compact and porous grains computed with the 3D SPH code is plotted on Fig.~\ref{sRphiCTTS}. Here as well, porous grains experience a quicker growth than compact grains. In 8090~yr, in the inner disc, the largest porous grains have planetesimal sizes and are in the Stokes drag regime while compact grains hardly reach one meter and stay in the Epstein regime. Furthermore, growth is very slow for compact dust beyond 100~au and grains cannot grow beyond 100~$\mu$m. On the contrary, porous dust can reach centimetre sizes up to 300~au and metre sizes interior to 100~au. The right column of Fig.~\ref{sRphiCTTS} shows that small grains are only slightly porous. As they grow, they become fluffier and fluffier and are compressed in the inner regions of the disc as seen in Fig.~\ref{zRphiCTTS} as well. 

\begin{figure}
\centering
\includegraphics[width=\columnwidth]{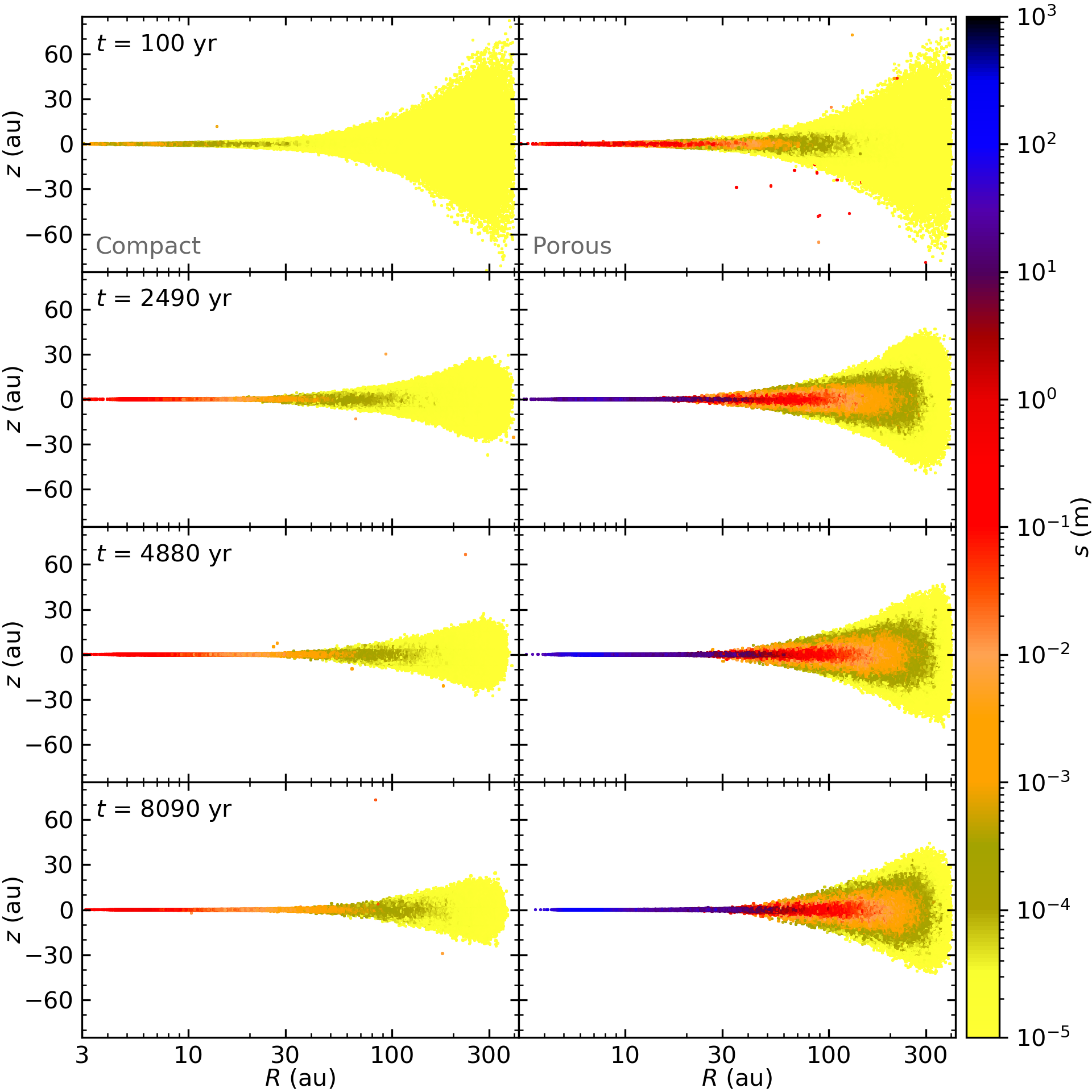}
\caption{Time evolution of the dust distribution in the CTTS disc obtained with the 3D SPH code. The colour bar represents the grain size. Porous grains (right) are less settled than compact dust (left). Four snapshots at 100, 2490, 4880 and 8090 yr are shown, from top to bottom.}    
\label{zRsCTTS}
\end{figure}

Fig.~\ref{zRsCTTS} shows the time evolution of the dust spatial distribution of compact and porous grains computed with the 3D SPH code. In both cases, dust is vertically size-sorted. Larger solids can be found close to the mid-plane while small grains are distributed over a larger scale height. In the porous case, the mid-plane is filled with $\sim100$~m solids from the inner edge out to $\sim50$~au. Two differences on dynamics can be also spotted out. After 8090~yr, the disc made of compact dust is approximately 20~au less radially extended than with porous grains. Moreover, this disc is also more settled compared to the one with fluffy dust. Indeed, porous grains can stay coupled with the gas phase ($\mathrm{St}\ll1$) at larger sizes than compact grains, they settle down and drift inward slightly less rapidly.

\subsection{Influence of monomer size}
\label{subsec3}

\begin{figure}
\centering
\includegraphics[width=\columnwidth]{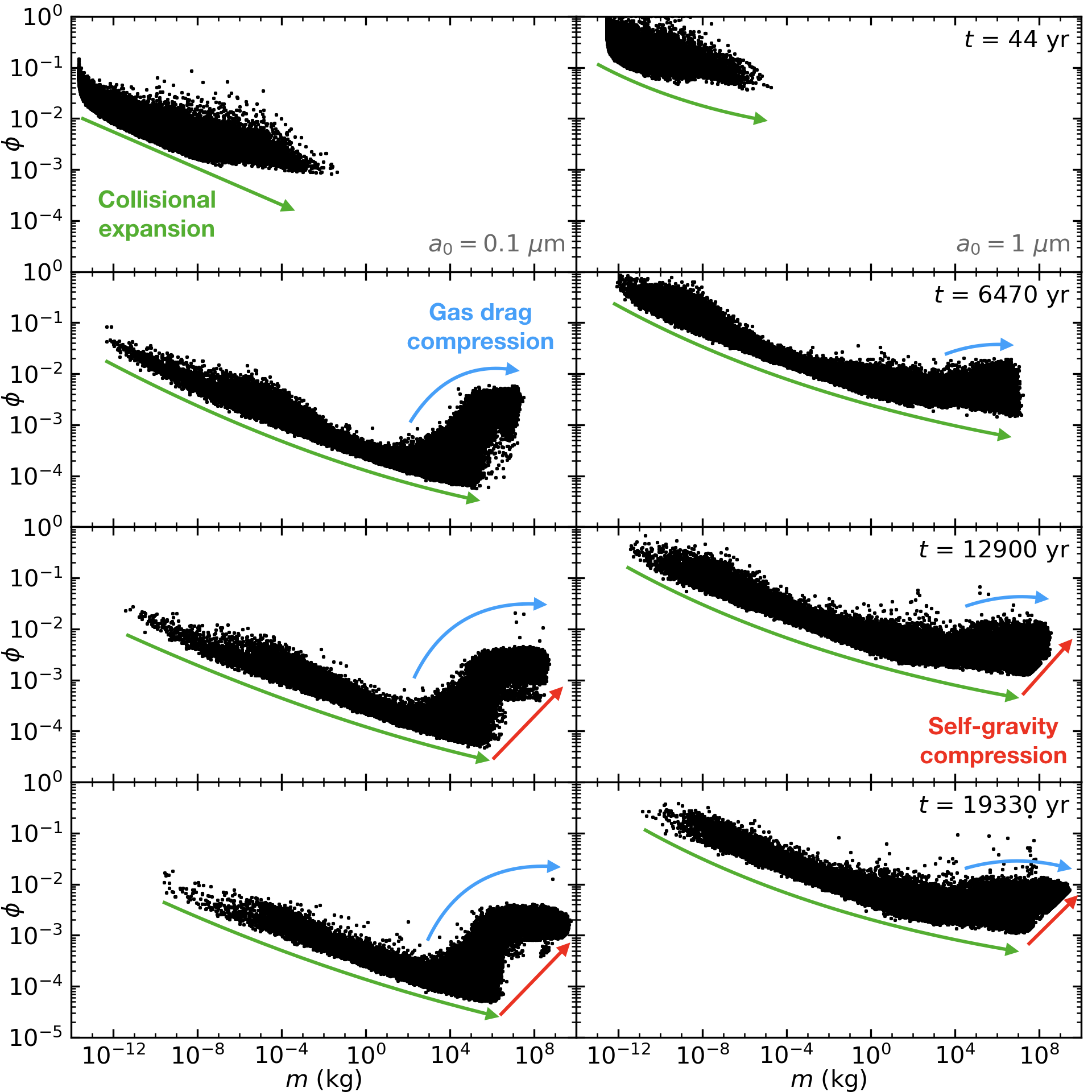}
\caption{Time evolution of the filling factor as a function of mass for grains in the Flat disc obtained with the 3D SPH code. Porous grains that are made of 0.1~$\mu$m monomers (left) can become more porous than grains formed with 1~$\mu$m monomers (right). However, the filling factor has a similar behaviour in both cases: it decreases thanks to the collisional expansion regime (green arrow) then increases because of the compression due to the gas drag (blue arrow) and finally, grains reach the self-gravity compression regime (red arrow). Four snapshots at 44, 6470, 12900 and 19330 yr are shown, from top to bottom.}    
\label{phimFlata0}
\end{figure}

Figure~\ref{phimFlata0} compares the time evolution of the filling factor during growth for grains made of monomers of different sizes $a_0=0.1$ and 1~$\mu$m in the Flat disc. The behaviours are similar to the CTTS disc (Fig. \ref{phimCTTS}). Initially, grains are in the collisional expansion regime then are compacted because of the gas drag. Grains reach similar masses in both cases, large enough to be compressed by their self-gravity as indicated by the red arrow. All these regimes depend on the monomer size and the filling factor increases when $a_\mathrm{0}$ increases, see equations~(\ref{eq:phiHS})--(\ref{phigrav}). Thus, 10-$\mu$m sized grains composed of 1-$\mu$m monomers are more massive than those made of 0.1-$\mu$m monomers. One can observe that gas drag compression is almost as efficient in both cases, i.e. it compresses grains to similar filling factors. However, since grains made of large monomers are already more compact, their change in filling factor is smaller during that phase. In the same way, self-gravity starts to compact the dust at a larger mass. 

\begin{figure}
\centering
\includegraphics[width=\columnwidth]{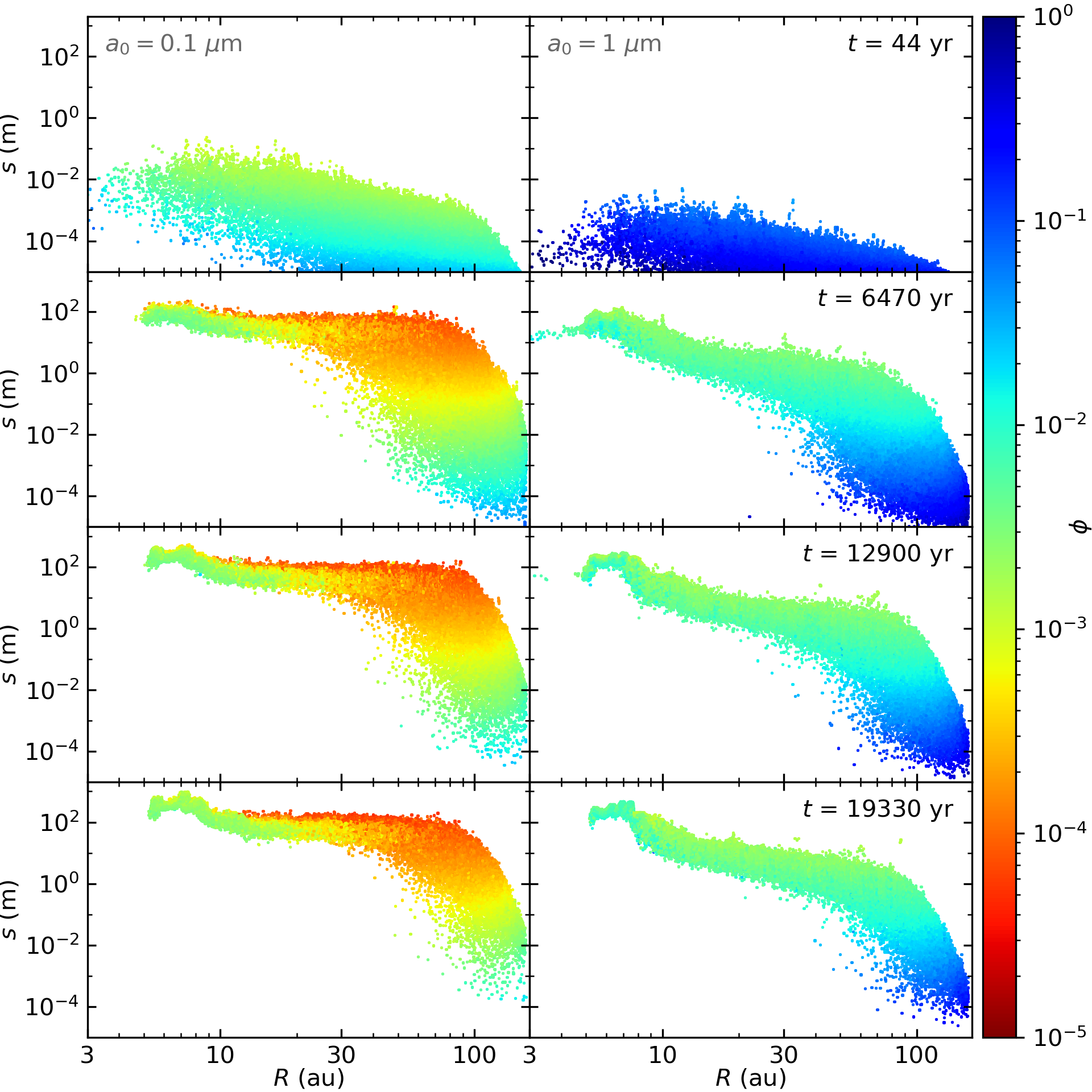}
\caption{Time evolution of the radial grain size distribution in the Flat disc obtained with the 3D SPH code. Aggregates made of 0.1-$\mu$m monomers (left) can reach larger sizes all over the disc than dust made of 1-$\mu$m monomers (right). The colour bar represents the filling factor. Four snapshots at 44, 6470, 12900 and 19330 yr are shown, from top to bottom.}
\label{sRphiFlata0}
\end{figure}

The radial size distribution is compared in Fig.~\ref{sRphiFlata0}. At first, the difference in filling factor does not affect much the radial extension of the disc, i.e. the radial drift is not influenced by the monomer size. The main impact of the porosity change is on growth. Since grains composed of large monomers are less porous, they grow less efficiently and produce less large grains. Dust made of 0.1-$\mu$m monomers can produce planetesimals in the innermost 100~au of the disc while it is only the case over a few au with of 1-$\mu$m monomers.

\section{Discussion}
\label{Sec4}

\subsection{Porosity in the Solar System}

\citet{2012ApJ...752..106O} studied the possibility of forming planetesimals from direct growth of porous dust. This idea was then completed by \citet{2013A&A...557L...4K} who added the compression due to gas drag and self-gravity of massive solids. They have thus determined the evolution of the filling factor during growth but without drift and collective effects in a Minimum Mass Solar Nebula (MMSN) disc. With our continuous model, the values we obtain are generally in a good agreement with those from \citet{2013A&A...557L...4K}. The minimum value reached by the filling factor, $\sim 10^{-4}$, is however slightly larger than that of \citet{2013A&A...557L...4K} because our grains drift to regions where the gas density is larger, and hence gas drag compression stronger.

The Solar System still contains porous bodies today. Evidence has been found for low densities in asteroids, such as 253 Mathilde \citep{1999Natur.402..155H,1999Icar..140....3V}, or meteorites \citep[e.g.][]{2002M&PS...37..661B}.
This may even be a more widespread property, with the recent suggestion that the first known interstellar object to visit our system, 1I/`Oumuamua, is a fractal dust aggregate with an ultra-low density of $10^{-2}$~kg\,m$^{-3}$ \citep{2019ApJ...885L..41F}.
Spatial exploration and measurements on comets have shown that these objects made of ice and silicates are also porous. Their average density falls between 400 and 600~kg\,m$^{-3}$ \citep{2006ApJ...652.1768B,2011ARA&A..49..281A}. As an example, \citet{2015Sci...347a1044S} determined that the density of comet 67P/Churyumov-Gerasimenko is of the order of 470~kg\,m$^{-3}$. The \textit{Rosetta} probe orbiting comet 67P and its lander \textit{Philae} have provided a wealth of data on this body. In particular, they have shown that its surface is covered with solids whose size varies from tens of micrometers to tens of meters \citep{2017MNRAS.469S.755B}. Among them, millimiter-sized grains with a density $<1$~kg\,m$^{-3}$ and a filling factor $\phi \sim 10^{-3}$ were found with the GIADA instrument \citep{2015ApJ...802L..12F}. As can be seen on Fig.~\ref{sRphiCTTS}, our results show that grains with similar sizes and filling factors are present over large regions in the disc. While such low filling factors can seem surprising at first, they are compatible with the measurements made on comet 67P.

\subsection{The Stokes regime, a key point}
\label{sec:StokesKey}

One of the main theoretical problems for dust evolution is the radial-drift barrier, i.e.\ the rapid inwards drift of grains with $\mathrm{St} = 1$ leading to dust accretion onto the star. A solution to this problem is either to stop the grains from drifting or for them to grow from $\mathrm{St}\ll 1$ to $\mathrm{St}\gg 1$ (for which the drift is slow) in a time shorter than the drift timescale.

Moreover, \citet{2012A&A...537A..61L} have shown that if dust reaches $\mathrm{St}> 1$ in the Stokes drag regime, it remains in the disc if $q \leq 2/3$. Discs satisfying this condition represent 90$\%$ of discs observed by \citet{2005ApJ...631.1134A,2007ApJ...671.1800A}. The Stokes regime has two advantages: the Stokes number increases as grains get closer to the star (while it decreases in the Epstein regime) and varies as $s^{2}$ (while as $s$ in the Epstein regime), see equation~(\ref{Stokesnumber})\footnote{For a power-law disc, $\mathrm{St}\propto s\,R^p$ in the Epstein regime and $\mathrm{St}\propto s^2\,R^{(q-3)/2}$ in the Stokes regime.}. Thus, in the Stokes drag regime, dust can reach large Stokes numbers more easily. Physically, it means that the grain is more efficiently slowed down by the stronger gas drag in the inner parts of the disc. Consequently, reaching the Stokes drag regime with $\mathrm{St}\sim 1$ is a key point for grains to remain in most discs and thus survive the radial-drift barrier.

In order to understand the influence of porosity on the Stokes number and the drag regime, one needs to compare two characteristic sizes:
\begin{itemize}
    \item the optimal size, $s_{\mathrm{St}=1}$, for which the grain Stokes number reaches unity, i.e. the size corresponding to the fastest drift;
    \item $9\lambda/4$, the transition size between the Epstein and Stokes drag regimes.
\end{itemize}
In a power-law disc, the optimal size $s_{\mathrm{St}=1}\propto\phi^{-1} \, R^{-p}$ in the Epstein regime and $s_{\mathrm{St}=1}\propto\phi^{-1/2} \, R^{(3-q)/4}$ in the Stokes regime. $9\lambda/4$ does not depend on the filling factor and varies as $R^{p+(3-q)/2}$. If $s_{\mathrm{St}=1} < 9\lambda/4$ (resp. $s_{\mathrm{St}=1} > 9\lambda/4$), grains reach the maximal drift velocity in the Epstein (resp. Stokes) regime.

\begin{figure}
\centering
\includegraphics[width=\columnwidth]{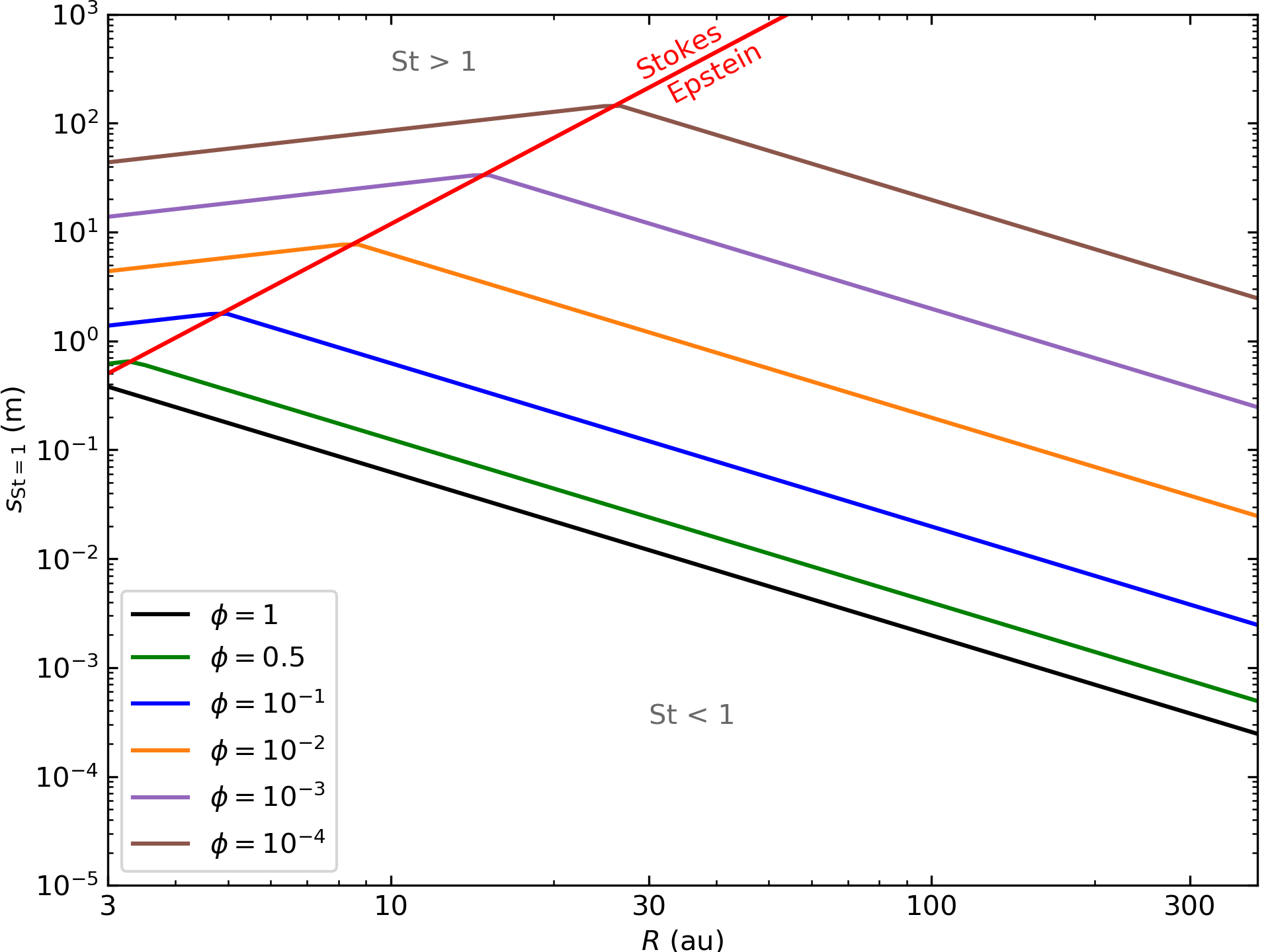}
\caption{Optimal size, i.e. size for which $\mathrm{St}=1$, as a function of the distance from the star in the CTTS disc for different filling factor values. For a given filling factor, below the line, grains have $\mathrm{St}< 1$ and above it $\mathrm{St}> 1$. The red line represents the limit between the Epstein and Stokes regimes for $s = 9\lambda/4$. Above this line, grains are in the Stokes regime and $\mathrm{St}\propto s^2$ while below, they are in the Epstein regime and $\mathrm{St}\propto s$.} 
\label{sSt1R}
\end{figure}

Figure~\ref{sSt1R} shows the optimal size for different filling factors and the transition between both drag regimes. For compact grains, the transition between the Epstein and Stokes regimes occurs necessarily with $\mathrm{St}> 1$. It means that compact grains experience a maximal inwards drift (and accretion onto the star) without the chance to transition to the Stokes regime. On the contrary, for grains with a filling factor lower than 0.5, there exists a region in the inner disc where porous grains are in the Stokes drag regime with $\mathrm{St}\leq 1$, which is more and more extended as grains become fluffier and fluffier. Thus, the Stokes number of those grains can increase rapidly during their growth, allowing them to decouple from the gas and survive the radial-drift barrier.

\subsection{Porosity and growth}

As discussed previously, porosity can have an impact on dust spatial evolution and growth. According to Figs.~\ref{sRCTTS} and \ref{sRphiCTTS}, compact and porous grains have a qualitatively similar growth behaviour in three stages. The optimal size for compact grains is about 0.1~mm. Since the optimal size depends on the grain filling factor, porous grains can reach larger sizes before starting to drift rapidly. For instance, porous grains with $\phi = 10^{-3}$, order of magnitude of the average filling factor over the disc (Fig.~\ref{sRphiCTTS}), have an optimal size $10^3$ times larger than for compact grains. Thus, they can grow up to $\sim10$~cm before drifting significantly. This average ratio of $10^3$ for porous grains sizes can be seen in Figs.~\ref{sRCTTS} and \ref{sRphiCTTS}. Generally, porous dust is characterised by a more efficient growth during the three steps described previously as their collision cross-section is larger than for compact grains. Consequently, porosity helps grains to reach larger sizes and transition to the Stokes drag regime in the inner disc, as seen in Figs.~\ref{sRCTTS} and \ref{sRphiCTTS}. We discussed in Section~\ref{sec:StokesKey} on how the Stokes drag regime is important for the dust survival in the disc.

Porosity also allows to form 100 m -- 1 km planetesimals in the mid-plane in the innermost 100~au, as shown by Figs.~\ref{sRCTTS} and \ref{zRsCTTS}, in less than $10^4$~yr. Even if our disc model is different from the one \citet{2016A&A...586A..20K} use, we find drift and growth timescales for both compact and porous grains similar to theirs. Those planetesimals are thought to be building blocks to form giant planet cores and our results can be used as a starting point for simulations of giant planet formation \citep{2016ApJ...817..105K,2018ApJ...862..127K}.

\subsection{On drift and the importance of collective effects}

\begin{figure}
\centering
\includegraphics[width=\columnwidth]{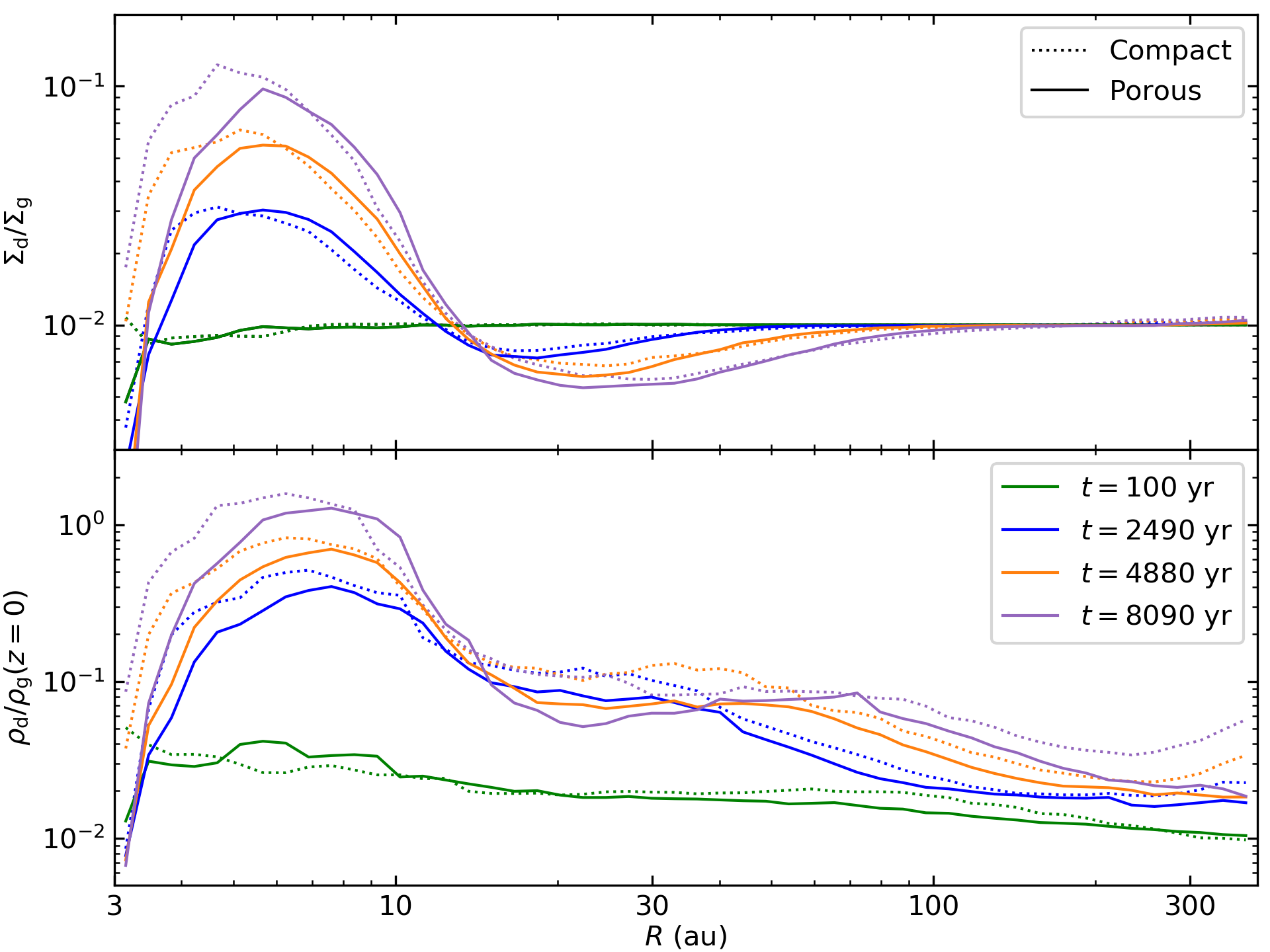}
\caption{Azimuthally-averaged radial profiles of the dust-to-gas ratio: vertically integrated (top) and in the mid-plane (bottom) for different time steps obtained with a 3D SPH simulation of the CTTS disc. Porous grains are represented by solid lines and compact grains by dotted lines. In the PACED code, the dust-to-gas ratio has a constant value of $10^{-2}$.}
\label{SigmaratioCTTS}
\end{figure}

We can spot a different behaviour for compact dust in the very inner disc. With the PACED code, compact dust is accreted onto the star without growing larger than the decimetre (Fig.~\ref{sRCTTS}), while we find in 3D SPH simulations that compact solids can remain in the very inner disc where they can reach metre sizes (Fig.~\ref{sRphiCTTS}). This phenomenon can be explained by the different calculation (and influence) of the local dust-to-gas ratio $\epsilon$ in our two codes. Indeed, \citet{1986Icar...67..375N} have shown (via equation~(\ref{vdre})) that dust drift is slowed down when the local dust-to-gas ratio increases thanks to dust collective effects. Moreover, the PACED code uses a constant $\epsilon$ of $10^{-2}$ while in the 3D SPH code, this quantity is directly calculated using the local dust and gas densities and so may vary. Figure~\ref{SigmaratioCTTS} reports the value of the dust-to-gas ratio at several times of the simulation with our 3D SPH code. Initially, $\epsilon = 10^{-2}$ in the whole disc as in the PACED code but as grains grow, they drift inwards and the dust-to-gas ratio increases in the innermost few au where its vertically integrated value can reach $10^{-1}$. In the mid-plane, it can even exceed unity thanks to vertical settling. Thus, the drift is slowed down and growth accelerated as the dust phase is denser. Those two effects help compact grains to reach meter sizes at the edge of the disc. However, they are not strong enough to allow them to reach the Stokes regime.

Such an increase in the dust-to-gas ratio in the inner disc was not seen for compact grains by \citet{2010A&A...513A..79B}, \citet{2012ApJ...752..106O}, or \citet{2016A&A...586A..20K}, whose framework and assumptions are very different from ours. These previous studies are all one-dimensional and compute the vertically-averaged dust evolution in the radial direction while keeping the gas distribution fixed (amounting to neglecting the back-reaction of dust on gas) and assuming vertical hydrostatic equilibrium for the gas and a prescription for the dust vertical scale-height as a function of turbulence. \citet{2012ApJ...752..106O} further assess the impact of back-reaction on the velocities of gas and dust and on collision velocities and find it to be negligible. However, it is not a full implementation of back-reaction as the background gas structure is still kept constant. For compact dust, all these studies found that grains grow until they reach $\mathrm{St}=1$ then drift inwards, producing a drift-limited size distribution and a flat dust-to-gas ratio profile. On the contrary, our simulations are three-dimensional and compute the self-consistent evolution of both gas and dust, including back-reaction of dust on gas. It has been shown that back-reaction, assisted by an increase of dust density close to the mid-plane due to vertical settling, is able to slow down the dust radial drift enough to allow grains to grow to sizes for which $\mathrm{St}> 1$, slow down their drift further, and pile up \citep[e.g.][]{2008A&A...487..265L,2015MNRAS.454L..36G,2017MNRAS.467.1984G}, something models neglecting back-reaction are unable to capture. The importance of back-reaction has further been shown by several authors \citep[e.g.][]{2017ApJ...844..142K,2017MNRAS.469.1932D,2018MNRAS.479.4187D}. The reader is referred to \citet{2017MNRAS.467.1984G} for a comparison of dust-to-gas ratio profiles with and without back-reaction in the case when dust fragmentation is included.

For porous dust, \citet{2012ApJ...752..106O} and \citet{2016A&A...586A..20K} found that grains grow rapidly, overcome the radial-drift barrier, then pile up, leading to an increase of the dust-to-gas ratio in the inner disc. This is exactly what we find in this study, see Section~\ref{subsec2} and Fig.~\ref{SigmaratioCTTS}. However, where \citet{2012ApJ...752..106O} saw an increase of their one-dimensional $\epsilon$ of a factor of several, our azimuthally-averaged, vertically-integraged $\epsilon$ is enhanced by one order of magnitude.
  Porous grains, similarly to compact grains, experience collective effects, which play an important role in setting their spatial distribution. Additionally, even though porosity can slighty slow down vertical settling and radial drift, its main effect remains a strong acceleration of growth (Section~\ref{subsec2}), which causes the transition to the Stokes drag regime. As a result, porous grains decouple from the gas and pile-up sooner in their evolution, therefore at larger distances from the star, than compact grains. This is reflected in the locations of the dust-to-gas ratio maxima in Fig.~\ref{SigmaratioCTTS}. The combination of collective effects and of the Stokes regime also results in a larger dust-to-gas ratio than when dust back-reaction is not included.

\section{Conclusion}
\label{Sec5}

The growth from sub-$\mu$m monomers to planetesimals is hampered by several barriers such as the radial-drift barrier. The dust needs to decouple from the gas in order to remain in the disc. In this work, we investigate how porosity can act on both grain growth and drift to overcome this barrier. It amounts to considering simultaneously, and thus link, the small (collisions) and large (disc) scales. To do so, we have developed a model of evolution of the filling factor depending on grain characteristics (bulk density, rolling energy, mass) and disc quantities (gas density, sound speed or temperature). Our model can be used for codes in which the disc evolution time step is different from that of individual collisions. We demonstrate that grains can remain in the disc in most cases if they grow enough to transition to the Stokes drag regime. Indeed, in this regime, dust can reach large Stokes numbers and decouple from the gas more easily. We have shown that compact grains do not grow quickly enough and stay in the Epstein regime. However, collective effects can slow them down enough for them to remain in the very inner regions of the CTTS disc. We find that the growth is accelerated for porous grains, allowing them to transition to the Stokes regime close to the star and survive the radial-drift barrier, both in the CTTS and Flat discs. Furthermore, our study has shown that porous millimetre grains have an average filling factor of about $10^{-3}$, in good agreement with measurements made on comet 67P/Churyumov-Gerasimenko \citep{2015ApJ...802L..12F}. Finally, we find that small planetesimals can be formed by direct coagulation of porous dust in the innermost 100~au. This result provides a link with the formation of giant planet cores from planetesimals.

\section*{Acknowledgements}

We thank the anonymous referee for their careful reading and assessment of our work, which helped to improve this paper.
The authors acknowledge funding from ANR (Agence Nationale de la Recherche)
of France under contract number ANR-16-CE31-0013 (Planet-Forming-Disks).
This research was partially supported by the Programme National de
Physique Stellaire and the Programme National de Plan\'etologie of
CNRS/INSU, France. AJLG and JFG thank the LABEX Lyon Institute of Origins
(ANR-10-LABX-0066) of the Universit\'e de Lyon for its financial
support within the programme `Investissements d'Avenir'
(ANR-11-IDEX-0007) of the French government operated by the ANR.
Simulations were run at the Common Computing Facility (CCF) of LABEX LIO.
All figures were made with with the Python library \texttt{matplotlib} \citep{matplotlib}.



\bibliographystyle{mnras}
\bibliography{refs}



\appendix

\section{Model of porosity evolution during collisions}
\label{App1}

\citet{2008ApJ...684.1310S} have found that the filling factor after collision $\phi_\mathrm{f}$ is related to the filling factor $\phi_\mathrm{i}$ and some other quantities such as the kinetic energy $E_\mathrm{kin}$ before collision by
\begin{equation}
\label{phioku}
\phi_\mathrm{f} = \left\{
    \begin{array}{l}
       2\phi_\mathrm{i} \left[2.99^{5/6} + (2 - 2.99^{5/6}) \displaystyle\frac{E_\mathrm{kin}}{3\,b\,E_\mathrm{roll}} \right]^{-6/5} \,, \\[3ex]
       \mbox{for} \: E_\mathrm{kin} \le 3\,b\,E_\mathrm{roll} \: \mbox{(hit and stick regime)}\\
 \\
      \displaystyle\frac{2m_\mathrm{i}}{\rho_\mathrm{s}} \left[\frac{(3/5)^5 \left(E_\mathrm{kin} - 3\,b\,E_\mathrm{roll}\right)}{N_\mathrm{tot}^5 \,b\,E_\mathrm{roll}\,V_\mathrm{0}^{10/3}} + \left(2 V_\mathrm{i}^{5/6}\right)^{-4} \right]^{3/10} \,, \\[4ex]
       \mbox{for} \: E_\mathrm{kin} \ge 3\,b\,E_\mathrm{roll} \: \mbox{(internal restructuring regime)}\\
    \end{array}
\right.
\end{equation}
where $m_\mathrm{i}$ is the mass of colliding grains and $V_\mathrm{i}$ their volume. $N_\mathrm{tot}$ is the total number of monomers involved in the collision, $V_\mathrm{0}$ is the volume of a monomer, defined as a compact sphere of radius $a_\mathrm{0}$ and $b$ is a numerical factor taken as 0.15 \citep{2012ApJ...752..106O}.

The rolling energy is given by \citet{1997ApJ...480..647D} as
\begin{equation}
E_\mathrm{roll}=6\pi^2\,\gamma\,a_0\,\xi_\mathrm{crit},
\end{equation}
where $\gamma$ is the surface energy of the material. They find that $\xi_\mathrm{crit}$, the critical rolling distance of one monomer on another for energy dissipation, is of the same order as the critical distance for monomer separation, given by \citet{1993ApJ...407..806C} as
\begin{equation}
\delta_\mathrm{c}=\left(\frac{27\pi^2\,\gamma^2\,a_0}{2\,{\cal E}^2}\right)^{1/3},
\end{equation}
where $\cal E$ is the Young modulus of the material. This leads to
\begin{equation}
\label{eq:Eroll}
E_\mathrm{roll}=\left(\frac{2916\,\pi^8\,\gamma^5\,a_0^4}{{\cal E}^2}\right)^{1/3},
\end{equation}
where we adopt for ice $\gamma=7.3\times10^{-2}$~J\,m$^{-2}$ and ${\cal E}=9.4$~GPa \citep{2014ApJ...783L..36Y}.

In this Appendix, we explain how we extend the \citet{2008ApJ...684.1310S} equation (\ref{phioku}) to make it continuous in Section~\ref{Apsub1} for the hit and stick regime (i.e. $E_\mathrm{kin} \leq 3 \, b \, E_\mathrm{roll}$) and in Section~\ref{Apsub2} for the internal restructuring regime (i.e. $E_\mathrm{kin} \geq 3 \, b \, E_\mathrm{roll}$). It is equivalent to express $\phi_\mathrm{f}$ as a power law of the grain mass $m$ and quantities of the disc. To do so, we chose to approximate $E_\mathrm{kin}$ as either very small or very large compared to $E_\mathrm{roll}$ in order to do a finite expansion. More detailed calculations leading to the equations presented in this Appendix can be found in \citet[in French]{garcia:tel-01977317}.

\subsection{Hit-and-stick regime}
\label{Apsub1}

If we consider $E_\mathrm{kin} \ll E_\mathrm{roll}$, equation (\ref{phioku}) becomes
\begin{equation}
\phi_\mathrm{f} = \frac{2}{2.99} \, \phi_\mathrm{i} \: .
\end{equation}
Thus, we see that the filling factor in the hit-and-stick regime evolves as a geometrical progression with a common ratio $2/2.99$. We can express the filling factor $\phi_\mathrm{f}$ of a grain with a mass $m$ as a function of the monomer filling factor $\phi_\mathrm{0}$  and $n$ the number of collisions to form that grain
\begin{equation}
\phi_\mathrm{f} = \phi_\mathrm{0} \left(\frac{2}{2.99}\right)^n .
\end{equation}
After one collision, the grain mass doubles. So after $n$ collisions from a monomer, $m = 2^n \, m_\mathrm{0}$. Consequently, the filling factor $\phi_\mathrm{f}$ in the hit-and-stick regime (thereafter renamed $\phi_{\mathrm{h\&s}}$) can be given as a function of $m$
\begin{equation}
\phi_{\mathrm{h\&s}} = \phi_\mathrm{0} \left(\frac{m}{m_\mathrm{0}}\right)^{\ln(2/2.99)/\ln(2)} .
\end{equation}
Since monomers are compact, $\phi_\mathrm{0} = 1$. Note that we no longer have the recursive aspect of equation~(\ref{phioku}). 

\subsection{Internal restructuring regime}
\label{Apsub2}

We consider here that $E_\mathrm{kin} \gg E_\mathrm{roll}$. Using $N_\mathrm{tot} = 2 m_\mathrm{i}/m_\mathrm{0}$ and $V_\mathrm{i}/V_\mathrm{0}$ = $1/\phi_\mathrm{i} \, (m_\mathrm{i}/m_\mathrm{0})$, equation~(\ref{phioku}) becomes
\begin{equation}
\label{phiokuapproxcol3}
\phi_\mathrm{f} = 2^{-1/5} \, \phi_\mathrm{i} \left[1 + \frac{(3/5)^5 \, E_\mathrm{kin}}{2 \, b \, E_\mathrm{roll}}\,\frac{1}{\phi_\mathrm{i}^{10/3}} \left(\frac{m_\mathrm{i}}{m_\mathrm{0}}\right)^{-5/3} \right]^{3/10} \:.
\end{equation}
However, $E_\mathrm{kin}$ depends on $\phi_\mathrm{i}$ trough $v_\mathrm{rel}^2$ and the Stokes number St. Nevertheless, $v_\mathrm{rel}^2$ does not vary linearly with the Stokes number. We consider St $\ll 1$ (resp. St $\gg 1$) in order to have $E_\mathrm{kin} \propto$ St (resp. St$^{-1}$). With this approximation, $\phi_\mathrm{f}$ is then related to $\phi_\mathrm{i}$ at a certain power. The power depends on the grain drag regime (Epstein or Stokes) and if its Stokes number is smaller or larger than 1. As we want to express $\phi_\mathrm{f}$ as $\phi_\mathrm{f} \propto m^k$ with $k \in \mathbb{R}$, we have for given disc quantities, $\phi_\mathrm{f} = 2^k \phi_\mathrm{i}$. We define $\beta = 2^{-k}$, and thus $\phi_\mathrm{i} = \beta \phi_\mathrm{f}$. The values of $\beta$ are different in Epstein and Stokes regimes and are taken to fit Eq. (\ref{phioku}) as discussed in Sections~\ref{Ap2.1} and \ref{Ap2.2}.  

\subsubsection{In the Epstein regime and St $<$ 1}
\label{Ap2.1}

In the Epstein regime with St $<$ 1, the filling factor $\phi_\mathrm{f}$ (thereafter renamed $\phi_\mathrm{Ep-St<1}$) of a grain with mass $m$ can be expressed as 
\begin{align}
&\phi_\mathrm{Ep-St<1} = \left(2^{1/5} - \beta_\mathrm{Ep}\right)^{-3/8}  \, \beta_\mathrm{Ep}^{-5/8} \, 2^{1/8} \nonumber \\
&\quad\times \left(\frac{3}{10} \frac{(3/5)^5 \, 2^{3/2} \, \mathrm{Ro}\, \alpha \, m_\mathrm{0} \, c_\mathrm{g}\, \rho_\mathrm{s} \, a_\mathrm{0} \, \Omega_\mathrm{K}}{8 \,\rho_\mathrm{g}\,b\,E_\mathrm{roll}}\right)^{3/8} \left(\frac{m}{m_\mathrm{0}}\right)^{-1/8} \,,
\end{align}
where $\beta_\mathrm{Ep}$ is the value of $\beta$ in the Epstein regime. As $k=-1/8$, $\beta_\mathrm{Ep} = 2^{1/8}$. 

\subsubsection{In the Stokes regime and St $<$ 1}
\label{Ap2.2}

In the Stokes regime with St $<$ 1, the filling factor $\phi_\mathrm{f}$ (thereafter renamed $\phi_\mathrm{St-St<1}$) of a grain with mass $m$ can be expressed as 
\begin{align}
\label{phiStSt<12}
&\phi_\mathrm{St-St<1} = \left(2^{1/5} - \beta_\mathrm{St}\right)^{-1/3} \, \beta_\mathrm{St}^{-2/3} \nonumber \\
&\quad\times \left(\frac{3}{10} \frac{(3/5)^5 \, 2^{3/2} \, \mathrm{Ro}\, \alpha \, m_\mathrm{0} \, c_\mathrm{g}^2\, \rho_\mathrm{s} \, a_\mathrm{0}^2 \, \Omega_\mathrm{K}}{36 \, \mu_\mathrm{g} \, b \, E_\mathrm{roll}}\right)^{1/3} \,,
\end{align}
where $\beta_\mathrm{St}$ is the value of $\beta$ in the Stokes regime. Since $\phi_\mathrm{St-St<1}$ does not depend on the mass, $\beta_\mathrm{St} = 1$. 

\subsubsection{In the Epstein and Stokes regimes and St $>$ 1}

When the Stokes number St becomes larger than unity, the right term in the bracket on equation~(\ref{phiokuapproxcol3}) can be negligible compared to 1. If $\mathrm{St} = 1$ is reached for a mass $M_\mathrm{4}$ (respectively $M_\mathrm{5}$) in the Epstein (resp. Stokes), the filling factor in this regime $\phi_\mathrm{Ep-St>1}$ (resp. $\phi_\mathrm{St-St>1}$) is expressed as
\begin{equation}
\label{phiEpSt>1}
\phi_\mathrm{Ep-St>1} = \phi_\mathrm{Ep-St<1}(M_\mathrm{4}) \left(\frac{m}{M_\mathrm{4}}\right)^{-1/5} \,,
\end{equation}
\begin{equation}
\label{phiStSt>1}
\phi_\mathrm{St-St>1} = \phi_\mathrm{St-St<1}(M_\mathrm{5}) \left(\frac{m}{M_\mathrm{5}}\right)^{-1/5} \,,
\end{equation}
where $M_\mathrm{4}$ and $M_\mathrm{5}$ are respectively given by equations~(\ref{M4}) and (\ref{M5}).

\subsection{Transition masses}
\label{Apsub3}


\begin{table}
\centering
\caption{Definitions of the transition masses.}
\label{Tableau3}
\begin{tabular}{| c || c || c |}
\hline
Transition masses & Transition \\
\hline
\hline
$M_\mathrm{1}$  & beginning of the collisional compression\\
& in the Epstein regime with St < 1 \\
\hline
$M_\mathrm{2}$  & beginning of the collisional compression\\
& in the Stokes regime with St < 1 \\
\hline
$M_\mathrm{3}$ & transition from the Epstein regime \\
 & to the Stokes regime with St < 1 \\
\hline
$M_\mathrm{4}$ & transition from St < 1 to St > 1 \\
 & in the Epstein regime \\
\hline
$M_\mathrm{5}$ & transition from St < 1 to St > 1 \\
 & in the Stokes regime \\
\hline
\end{tabular}
\end{table}

In order to know in which regime the grain is, we compare its mass $m$ with the limit masses in every regime with the algorithm. The names of those transition masses are reported in Table~\ref{Tableau3}. The transition masses $M_\mathrm{1}$ to $M_\mathrm{5}$ are given by
\begin{align}
\label{M1}
& \frac{M_\mathrm{1}}{m_\mathrm{0}} = \left[\left(2^{1/5} - \beta_\mathrm{Ep}\right)^{-3/8} \, \beta_\mathrm{Ep}^{-5/8} \, 2^{1/8} \right. \nonumber \\
& \quad\times \left. \left(\frac{3}{10} \frac{(3/5)^5 \, 2^{3/2} \, \mathrm{Ro}\, \alpha \, m_\mathrm{0} \, c_\mathrm{g}\, \rho_\mathrm{s} \, a_\mathrm{0} \, \Omega_\mathrm{K}}{8 \, \rho_\mathrm{g}\,b\, E_\mathrm{roll}}\right)^{3/8} \right]^\frac{\scriptstyle1}{\textstyle\frac{1}{8} + \frac{\ln(2/2.99)}{\ln(2)}} ,
\end{align}
\begin{align}
\label{M2}
& \frac{M_\mathrm{2}}{m_\mathrm{0}} = \left[\left(2^{1/5} - \beta_\mathrm{St}\right)^{-1/3}\,\beta_\mathrm{St}^{-2/3}\right.\nonumber \\
& \quad\times \left. \left(\frac{3}{10} \frac{(3/5)^5 \, 2^{3/2} \, \mathrm{Ro}\, \alpha \, m_\mathrm{0} \, c_\mathrm{g}^2 \, \rho_\mathrm{s} \, a_\mathrm{0}^2 \, \Omega_\mathrm{K}}{36 \, \mu_\mathrm{g} \, b \, E_\mathrm{roll}}\right)^{1/3}\right]^\frac{\scriptstyle\ln(2)}{\scriptstyle\ln(2/2.99)},
\end{align}
\begin{align}
\label{M3}
&\frac{M_\mathrm{3}}{m_\mathrm{0}} = \left(2^{1/5} - \beta_\mathrm{St}\right)^{8/3} \left(2^{1/5} - \beta_\mathrm{Ep}\right)^{-3} \,\beta_\mathrm{St}^{16/3}\,\beta_\mathrm{Ep}^{-5} \, 36^{8/3} \, 8^{-3} \nonumber \\
& \quad\times \left(\frac{3}{10} \frac{(3/5)^5 \, 2^{3/2} \, \mathrm{Ro}\, \alpha \, m_\mathrm{0}  \, \rho_\mathrm{s} \, \Omega_\mathrm{K}}{b \, E_\mathrm{roll}}\right)^{1/3} 
 c_\mathrm{g}^{-7/3} \, a_\mathrm{0}^{-7/3} \, \rho_\mathrm{g}^{-3} \, \mu_\mathrm{g}^{8/3} ,
\end{align}
\begin{align}
\label{M4}
&\frac{M_\mathrm{4}}{m_\mathrm{0}} = \left(\frac{\rho_\mathrm{g} \, c_\mathrm{g}}{\Omega_\mathrm{K} \, \rho_\mathrm{s} \,a_\mathrm{0}}\right)^4 2^{-1/3}\,\beta_\mathrm{Ep}^{5/3}\left(2^{1/5}-\beta_\mathrm{Ep}\right) \nonumber\\
&\quad \times \left(\frac{3}{10} \frac{(3/5)^5 \, 2^{3/2} \, \mathrm{Ro}\, \alpha \, m_\mathrm{0} \, c_\mathrm{g}\, \rho_\mathrm{s} \, a_\mathrm{0} \, \Omega_\mathrm{K}}{8\,\rho_\mathrm{g}\,b\,E_\mathrm{roll}}\right)^{-1},
\end{align}
\textsc{ }\\
\begin{align}
\label{M5}
&\frac{M_\mathrm{5}}{m_\mathrm{0}} = \left(\frac{9 \, \mu_\mathrm{g}}{2 \, \Omega_\mathrm{K} \, \rho_\mathrm{s} \, a_\mathrm{0}^2}\right)^{3/2} \beta_\mathrm{St}^{1/3} \left(2^{1/5} - \beta_\mathrm{St}\right)^{1/6} \nonumber\\
&\quad \times \left(\frac{3}{10} \frac{(3/5)^5 \, 2^{3/2} \, \mathrm{Ro}\, \alpha \, m_\mathrm{0} \, c_\mathrm{g}^2 \, \rho_\mathrm{s} \, a_\mathrm{0}^2 \, \Omega_\mathrm{K}}{36\,\mu_\mathrm{g}\,b\, E_\mathrm{roll}}\right)^{-1/6}.
\end{align}
Note that those expressions depend only on the distance from the star $R$ in the case of a power-law disc model. 

\begin{figure}
\centering
\includegraphics[width=\columnwidth]{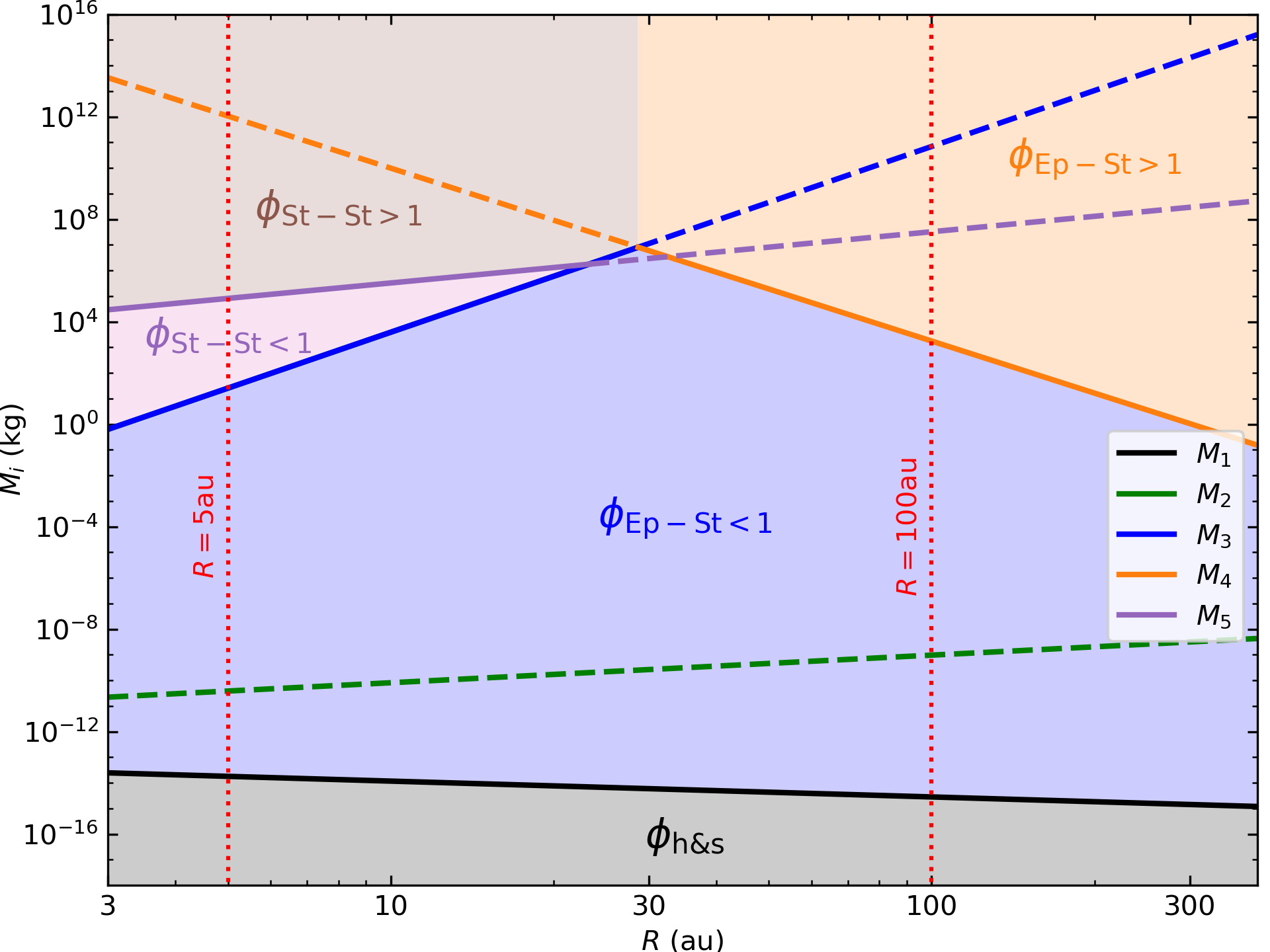}
\caption{Different regimes of evolution of the filling factor encountered during collisions according to grain mass and distance from the star in the CTTS disc. Straight lines represent transitions between regimes. Dashed lines are used when transitions are no longer operating. The two vertical dotted red lines illustrate the regimes crossed by grains at 5 and 100 AU.}
\label{Regimesphicol}
\end{figure}

\begin{figure}
\centering
\includegraphics[width=\columnwidth]{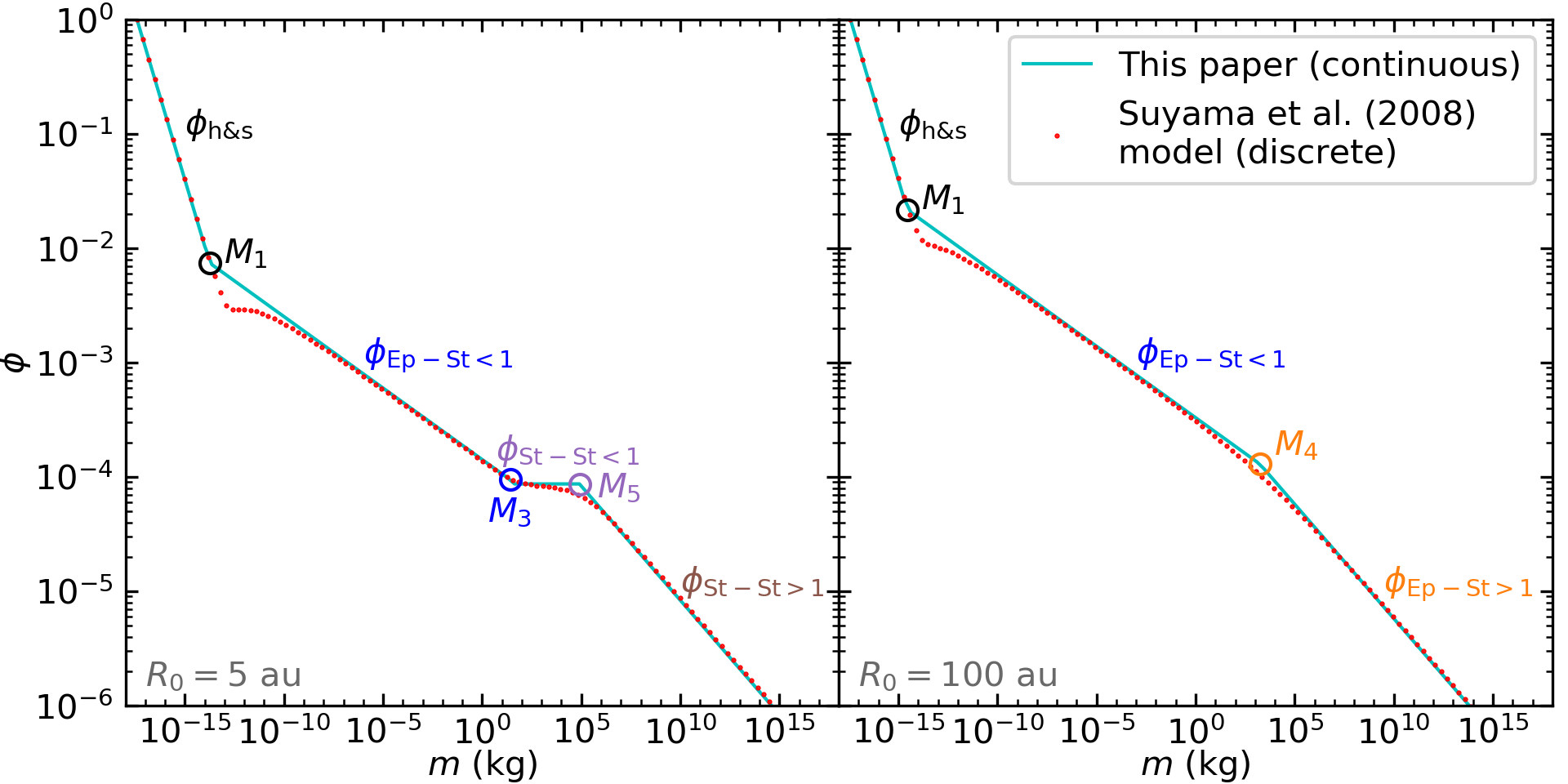}
\caption{Comparison of our model of filling factor evolution during collisions (straight cyan lines) with the \citet{2008ApJ...684.1310S} discrete model (red dots) for grains evolving at fixed positions of 5~au (left) and 100~au (right) in the CTTS disc.}
\label{Comparaison}
\end{figure}

The different transition masses are plotted in Fig.~\ref{Regimesphicol} for the CTTS disc. As shown in Figs.~\ref{Regimesphicol} and \ref{Comparaison}, all the dust begins to grow in the hit-and-stick regime before getting compacted by collisions in the Epstein drag regime. However, two different cases can be highlighted: grains in the first tens of AU transition to the Stokes regime as they grow while further in the disc, they keep growing in the Epstein regime. Moreover, Fig.~\ref{Comparaison} exhibits a comparison between the \citet{2008ApJ...684.1310S} discrete model and our continuous model. Both models mostly give similar results, allowing the simpler implementation of the continuous model in hydrodynamical codes. Our model deviates from that of \citet{2008ApJ...684.1310S} in two situations: for $E_\mathrm{cin} \sim E_\mathrm{roll}$ and $\mathrm{St}\sim 1$. In both cases, the filling factor value we obtain is slightly higher than obtained with the discrete model but this difference is not significant for grains growth and dynamics and can be neglected.

\subsection{Algorithm}
\label{ApAlgo}

The algorithm allowing to compute the value of the filling factor after collisions $\phi_\mathrm{col}$ is detailed in Algorithm~\ref{algorithmphicol}.

\begin{algorithm}
\caption{Calculation of $\phi_\mathrm{col}$}
\label{algorithmphicol}
\begin{algorithmic} 
\IF{$M_\mathrm{2} < M_\mathrm{1}$}
\IF{$m < M_\mathrm{2}$}
\STATE $\phi_\mathrm{col} = \phi_\mathrm{h\&s}$
\ELSE
\IF{$m < M_\mathrm{5}$}
\STATE $\phi_\mathrm{col} = \phi_\mathrm{St-St<1}$
\ELSE 
\STATE $\phi_\mathrm{col} = \phi_\mathrm{St-St>1}$
\ENDIF
\ENDIF
\ELSE
\IF{$m < M_\mathrm{1}$}
\STATE $\phi_\mathrm{col} = \phi_\mathrm{h\&s}$
\ELSE
\IF{$M_\mathrm{4} > M_\mathrm{3}$}
\IF{$m < M_\mathrm{3}$}
\STATE $\phi_\mathrm{col} = \phi_\mathrm{Ep-St<1}$
\ELSE
\IF{$m < M_\mathrm{5}$}
\STATE $\phi_\mathrm{col} = \phi_\mathrm{St-St<1}$
\ELSE
\STATE $\phi_\mathrm{col} = \phi_\mathrm{St-St>1}$
\ENDIF
\ENDIF
\ELSE
\IF{$m < M_\mathrm{4}$}
\STATE $\phi_\mathrm{col} = \phi_\mathrm{Ep-St<1}$
\ELSE 
\STATE $\phi_\mathrm{col} = \phi_\mathrm{Ep-St>1}$
\ENDIF
\ENDIF
\ENDIF
\ENDIF
\end{algorithmic}
\end{algorithm}

\section{Porosity evolution of static grains in the full model}

\begin{figure}
\centering
\includegraphics[width=\columnwidth]{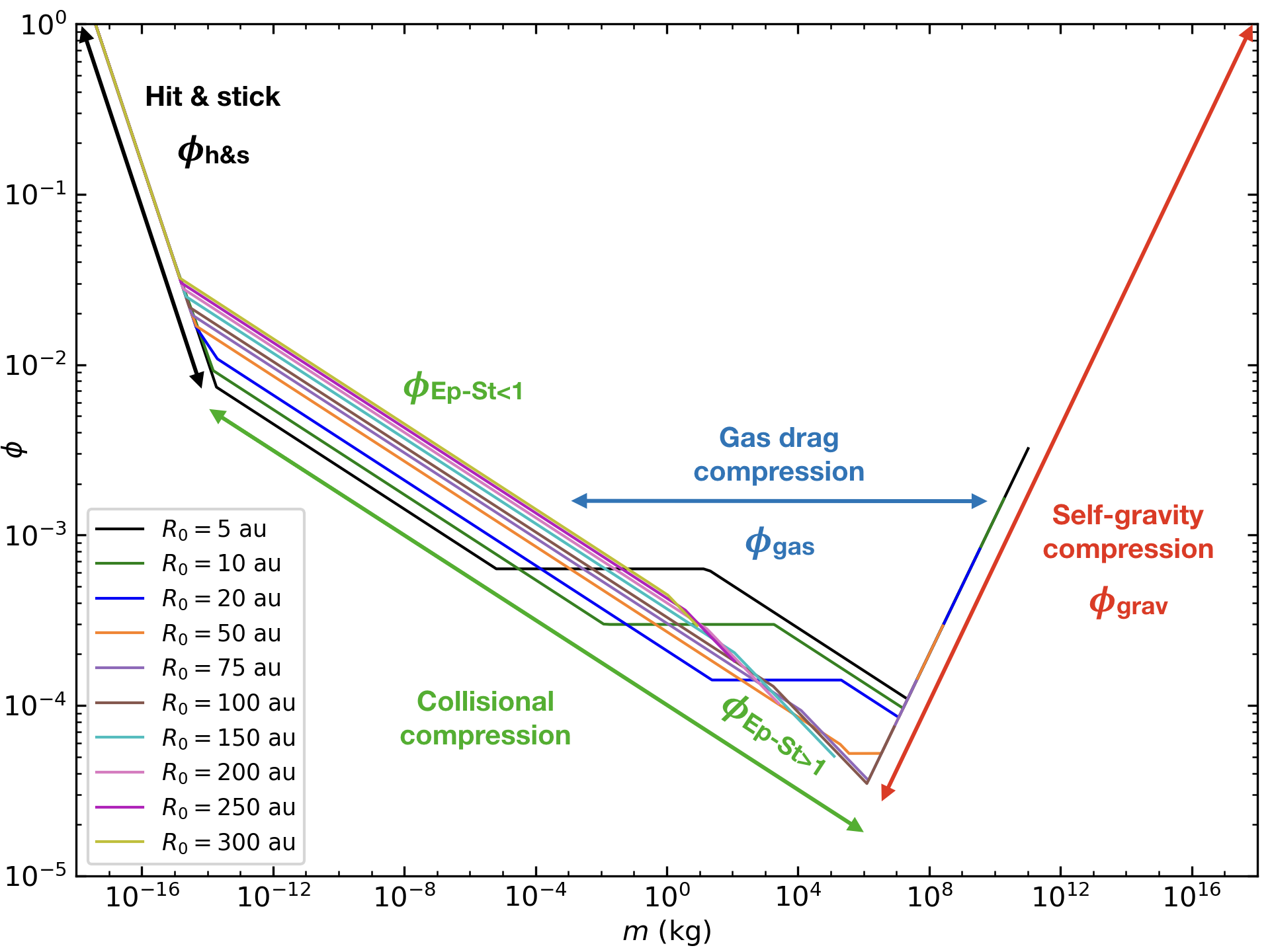}
\caption{Grain filling factor as a function of mass in the CTTS disc at different fixed positions $R_\mathrm{0}$ with the PACED code. Arrows show the different regimes encountered by the grains.}
\label{phimana2}
\end{figure}

Figure~\ref{phimana2} shows the evolution of the filling factor of static grains as they grow at different distances from the star as a function of their mass, computed with the PACED code. The different regimes in the collisional evolution (hit-and-stick and collisional compression) and static compression (gas drag and self-gravity) phases are identified. Figure~\ref{phimana2} is to be compared with Fig.~\ref{phimana}, showing the porosity evolution of radially drifting grains (see Section~\ref{subsec1}).


\bsp	
\label{lastpage}
\end{document}